\documentclass[fleqn,usenatbib,useAMS,usenatbib]{mnras}

\catcode`\==12        
\mathcode`\=="303D    

\usepackage{newtxtext,newtxmath} 
\usepackage{microtype}
\usepackage{graphicx}
\usepackage{xurl}
\usepackage{xspace}
\usepackage{amsmath}
\usepackage{booktabs}
\usepackage{threeparttablex}
\usepackage{longtable}
\usepackage{multirow}
\usepackage{tabularx}
\usepackage{placeins}
\usepackage{subcaption}
\usepackage{float}

\usepackage{siunitx}
\DeclareRobustCommand{\VAN}[3]{#2}
\let\VANthebibliography\thebibliography
\def\thebibliography{\DeclareRobustCommand{\VAN}[3]{##3}\VANthebibliography}

\usepackage{graphicx}	
\usepackage{amsmath}	
\usepackage{siunitx}
\title[TOI-4616]{TOI-4616\,b: a benchmark Earth-sized planet transiting a nearby M4 dwarf}

\author[F. Zong Lang et al.]{F.~Zong~Lang$^{1}$ 
~B. O. Demory$^{1}$,
~Y. G\'omez~Maqueo~Chew$^{2}$,
~Y.~Schmid$^{1}$,
M. ~Timmermans$^{8,3}$,
~F.J.~Pozuelos$^{26}$,
\newauthor
M.~Gillon$^{3}$,
~Artem~Y.~Burdanov$^{14}$,
~Benjamin V.~Rackham$^{14,18}$,
D.~Queloz$^{6,33}$,
K.G.~Stassun$^{7}$,
~Khalid~Barkaoui$^{15,3,14}$,
\newauthor
Amaury ~Triaud$^{3,13}$,
~J. de Wit$^{9}$,
~S ~Zúñiga-Fernández$^{12}$,
~A.~J.~Burgasser$^{22}$,
~Elsa Ducrot$^{10}$,
Madison ~G. ~Scott$^{8}$,
\newauthor
D.~Sebastian$^{9,10}$,
~A. Soubkiou$^{3}$,
~M.~Lendl$^{4}$,
I.~Plauchu-Frayn$^{5}$,
U.~Schroffenegger$^{1}$,
~Erik Meier V.$^{32}$,
~P.~Pedersen$^{6,33}$
\newauthor
A.~Khandelwal$^{2}$,
Roman~Gerasimov$^{24}$,
~C. ~Aganze$^{23}$,
~Chih-Chun~Hsu$^{25}$,
~J.M.~Jenkins$^{21}$,
Aishwarya R.~Iyer$^{27}$,
\newauthor
C. ~Watkins$^{11}$,
~C.~A.~Theissen$^{22}$,
~K.~A.~Collins$^{11}$,
~H.~P.~Osborn$^{28}$,
~A.~Shporer$^{18}$,
~Claudia Janô Muñoz$^{6}$,
\newauthor
Toshi Suganuma$^{30}$,
~Norio ~Narita$^{15,16,17}$,
~Akihiko Fukui$^{15,16}$,
F. Murgas$^{15,29}$,
J. de Leon$^{16}$,
Enric Pall\'{e}$^{15,29}$,
\newauthor
~Yasmin Davis$^{8}$
D. Kitzmann$^{1}$, 
Aishwarya R.~Iyer$^{27}$,
M. Pichardo Marcano$^{2}$,
M.J.~Hooton$^{6}$\\
$^{1}$Center for Space and Habitability, University of Bern, Gesellschaftsstrasse 6, 3012, Bern, Switzerland\\
$^{2}$Universidad Nacional Aut\'onoma de M\'exico, Instituto de Astronom\'ia, AP 70-264, CDMX 04510, M\'exico\\ 
$^{3}$Astrobiology Research Unit, Universit\'e de Li\'ege, All\'ee du 6 Ao\^ut 19C, B-4000 Li\'ege, Belgium\\ 
$^{4}$Observatoire de l’Universit\'e de Gen\'eve, Chemin des Maillettes 51, CH-1290 Versoix, Switzerland\\ 
$^{5}$Universidad Nacional Aut\'onoma de M\'exico, Instituto de Astronom\'ia, AP 106, Ensenada 22800, BC, M\'exico\\ 
$^{6}$Cavendish Laboratory, JJ Thomson Avenue, Cambridge, CB3 0HE, UK\\ 
$^{7}$Department of Physics \& Astronomy, Vanderbilt University\\ 
$^{8}$School of Physics and Astronomy, University of Birmingham, Edgbaston, Birmingham, B15 2TT, UK\\ 
$^{9}$Massachusetts Institute of Technology, Cambridge, MA, USA\\ 
$^{10}$LESIA Observatoire de Paris, Section de Meudon 5, place Jules Janssen 92195 MEUDON Cedex\\ 
$^{11}$Center for Astrophysics, Harvard \& Smithsonian, Observatory Building E, 60 Garden St, Cambridge, MA 02138\\ 
$^{12}$Département d'Astrophysique, ULiège, Belgique\\ 
$^{13}$University of Birmingham, Birmingham, UK\\ 
$^{14}$Department of Earth, Atmospheric and Planetary Science, Massachusetts Institute of Technology,77 Massachusetts Avenue,Cambridge, MA 02139, USA\\ 
$^{15}$Instituto de Astrof\'isica de Canarias (IAC), Calle V\'ia L\'actea s/n, 38200, La Laguna, Tenerife, Spain\\ 
$^{16}$Komaba Institute for Science, The University of Tokyo, 3-8-1 Komaba, Meguro, Tokyo 153-8902, Japan\\ 
$^{17}$Astrobiology Center, 2-21-1 Osawa, Mitaka, Tokyo 181-8588, Japan\\ 
$^{18}$Kavli Institute for Astrophysics and Space Research, Massachusetts Institute of Technology, Cambridge, MA, USA\\ 
$^{19}$ETH Zurich, Zurich, Switzerland.\\ 
$^{20}$University of Harvard, Harvard, USA.\\ 
$^{21}$NASA Ames Research Center, Moffett Field, CA 94035, USA\\ 
$^{22}$Department of Astronomy \& Astrophysics, UC San Diego, La Jolla, CA 92093, USA\\ 
$^{23}$Kavli Institute for Particle Astrophysics \& Cosmology, Stanford University, Stanford, CA 94305, USA\\ 
$^{24}$Department of Physics and Astronomy, University of Notre Dame, Nieuwland Science Hall, Notre Dame, IN 46556, USA\\ 
$^{25}$Center for Interdisciplinary Exploration and Research in Astrophysics (CIERA), Northwestern University, 1800 Sherman Ave, Evanston, IL, 60201, USA\\ 
$^{26}$Instituto de Astrof\'isica de Andaluc\'ia (IAA-CSIC), Glorieta de la Astronom\'ia s/n, 18008 Granada, Spain\\ 
$^{27}$NASA Postdoctoral Fellow, NASA Goddard Space Flight Center, 8800 Greenbelt Road, Greenbelt, MD 20771, USA\\
$^{28}$Center for Space \& Habitability, Physikalisches Institut, Universität Bern, Gesellschaftsstrasse 6, 3012 Bern, Switzerland\\ 
$^{31}$Inst. f. Teilchen- und Astrophysik, ETH Zürich, Wolfgang-Pauli-Strasse 27, 8093 Zürich, Switzerland\\ 
$^{29}$Departamento de Astrof\'isica, Universidad de La Laguna (ULL), E-38206 La Laguna, Tenerife, Spain\\
$^{30}$Department of Multi-Disciplinary Sciences, Graduate School of Arts and Sciences, The University of Tokyo, 3-8-1 Komaba, Meguro, Tokyo 153-8902, Japan 
$^{32}$Astrophysics, University of Oxford, Denys Wilkinson Building, Keble Road, Oxford, OX1 3RH, UK
$^{33}$Institute for Particle Physics and Astrophysics , ETH Z\"urich, Wolfgang-Pauli-Strasse 2, 8093 Z\"urich, Switzerland
}
\date{Accepted XXX. Received YYY; in original form ZZZ}
\pubyear{\the\year{}}
\begin{document}

\label{firstpage}
\pagerange{\pageref{firstpage}--\pageref{lastpage}}
\maketitle

\begin{abstract}

Rocky exoplanets are particularly abundant around M-type stars. Their small radii and low luminosities provide favourable conditions for detecting transiting terrestrial planets and probing their atmospheric properties.

We report the discovery and statistical validation of TOI-4616\,b, an Earth-sized planet transiting a nearby mid-M dwarf observed by TESS. We confirm the planetary nature of the signal and determine the system parameters by combining TESS photometry with ground-based multi-band transit observations, high-resolution imaging, and optical/near-infrared spectroscopy.

The host star lies at a distance of $28.10\pm0.07$\,pc and has $R_\star = 0.1889\pm0.0096\,R_\odot$, $M_\star = 0.1881\pm0.0094\,M_\odot$, and $T_{\rm eff}=3150\pm75$\,K. TOI-4616\,b has a radius of $1.22\,R_\oplus$ and an orbital period of 1.55\,d. The planet receives an incident flux of $\sim 40\,S_\oplus$, corresponding to an equilibrium temperature of $\sim 525$\,K. This places TOI-4616\,b in a regime intermediate between Earth-sized planets orbiting early M dwarfs and those around ultra-cool hosts.

Statistical validation with \texttt{TRICERATOPS}, supported by high-resolution imaging and chromatic transit constraints, yields a false-positive probability of $0.0135$ below the recommended threshold of $0.015$, confirming TOI-4616\,b as a validated planet. Owing to its proximity to Earth, well-constrained stellar properties, and extensive multi-band follow-up, TOI-4616\,b constitutes a valuable benchmark system for comparative studies of terrestrial planets around mid-M dwarfs and for future atmospheric investigations.

\end{abstract}

\begin{keywords}
planets and satellites: detection --
planets and satellites: fundamental parameters --
stars: low-mass --
stars: individual: TOI-4616 --
techniques: photometric
\end{keywords}



\section{Introduction}
The detection of terrestrial planets around mid-M dwarfs provides an empirical window into planet formation that cannot be gleaned from Solar System architectures alone \citep{chabrier2014giant, chachan2023small}. The relatively small radii and low luminosities of these stars enhance transit detectability and increase the amplitude of atmospheric spectral signatures \citep{dressing2013occurrence, Kreidberg2015, gratton2025discovery}. Systems hosting short period terrestrial planets, such as the M-dwarf system GJ\,1132 and its transiting planet GJ\,1132\,b illustrate how short-period, Earth-sized planets around cool hosts can serve as laboratories for atmospheric escape, photochemistry, and interior–atmosphere coupling \citep{libby2022featureless, palle2025exploring}.

Over the past decade, the discovery and characterization of exoplanets has advanced rapidly, driven in large part by NASA's Transiting Exoplanet Survey Satellite (TESS; \citealt{ricker2015transiting}). 
Among its detections is the nearby mid-M dwarf LP~466-156 (TIC~258796169; hereafter TOI-4616) with radius $0.1889\,R_{\odot}$ and mass $0.1881\,M_{\odot}$, around which a transiting planet candidate TOI-4616\,b (also known as LP~466-156\,b) with radius $1.22\,R_\oplus$ and orbital period of 1.55 days was identified. While the existence of the candidate signal is well established from TESS photometry, its validation and detailed characterization require coordinated ground-based follow-up, spectroscopic analysis, and statistical vetting.

The statistical vetting of transiting exoplanet candidates has become a mature subfield in its own right. Comprehensive validation frameworks now combine stellar population synthesis, Galactic structure modeling, and priors for astrophysical false-positive scenarios to assess the probability that a candidate is genuinely planetary. Early validation efforts \citep{prvsa2011kepler, kochoska2017gaia, bray2025statistical}, developed for the Kepler mission, established the methodology for quantifying blend frequencies and background eclipsing binary rates. More recent tools such as \texttt{VESPA} \citep{mistry2023vatest, mantovan2022validation} and \texttt{TRICERATOPS} \citep{giacalone2020vetting, giacalone2023discovery} have expanded these techniques by incorporating multi-band transit depth constraints, high-resolution imaging limits, Gaia astrometry, and instrument-specific dilution corrections. In particular, \texttt{TRICERATOPS}, optimized for TESS targets, employs a hierarchical Bayesian framework to compute vetting probabilities and is especially well suited for M-dwarf systems where unresolved companions can complicate the interpretation of shallow transits.

A crucial component of this validation process is the exclusion of astrophysical false positives through complementary observational diagnostics. Multi-band transit photometry provides a decisive test: For genuine planetary transits, the depth is expected to be largely wavelength-independent across broad optical and near-infrared bands, aside from small variations introduced by atmospheric transmission features. In contrast, astrophysical false positives such as blended eclipsing binaries often produce strong chromatic transit depth variations \citep{mislis2010multi, neckel1995solar}. Modern multi-band facilities routinely achieve the precision required to measure chromatic variations in transit depth \citep{Narita:2020}, enabling the exclusion of blended eclipsing binaries via chromatic-depth analysis \citep{pelaez2024validation, Ghachoui2024}. Complementary high-resolution imaging further constrains the presence of bound or line-of-sight companions that may contaminate the photometry \citep{ziegler2021soar, howell2021speckle, lester2021speckle}, thereby strengthening the statistical assessment of planetary authenticity.

Ground-based facilities optimized for faint red hosts, such as the \textsc{SPECULOOS} telescopes \citep{sebastian2021speculoos, sebastian2020development, delrez2018speculoos}, the \textsc{SAINT-EX} 1-m telescope in San Pedro Mártir \citep{demory2020super}, and \textsc{MuSCAT2} at Teide Observatory \citep{2019JATIS...5a5001N} play an essential role in refining transit parameters, measuring chromatic effects, and strengthening statistical validation. When combined with spectroscopic characterization and Spectral Energy Distribution (SED) fitting, these observations enable a self-consistent determination of host-star properties and therefore of planetary parameters.

Owing to its proximity, well-constrained stellar properties, and the availability of extensive multi-instrument photometry and spectroscopy, TOI-4616 represents a valuable benchmark system for studies of terrestrial planets orbiting mid-M dwarfs. In this work we therefore present not only the confirmation of the planet but a comprehensive characterization of the system, with the goal of establishing TOI-4616 as a reference target for comparative studies of small planets around late-type stars with well-characterized host and extensive follow-up observations.

We present a comprehensive analysis of the TOI-4616 system that integrates the TESS discovery data (Sect.~\ref{sec:tess}) with ground-based multi-instrument photometry (Sect.~\ref{subsec:Ground_based_phot}), high-resolution imaging (Sect.~\ref{sec:High_res_imaging}), and optical/near-infrared spectroscopy (Sect.~\ref{sec:stell_spect}) used for stellar characterization (Sect.~\ref{sec:Stell_charact}). We adopt a uniform reduction and modeling framework, performing a joint transit analysis to derive the planet’s physical parameters (Sect.~\ref{sec:modeling}) and compute false-positive probabilities under a broad set of astrophysical scenarios (Sect.~\ref{sec:validation}). 

We conclude with a discussion of the broader implications of this system in Sect.~\ref{sec:disc}.


   \begin{figure}
      \centering
      \includegraphics[width=\columnwidth]{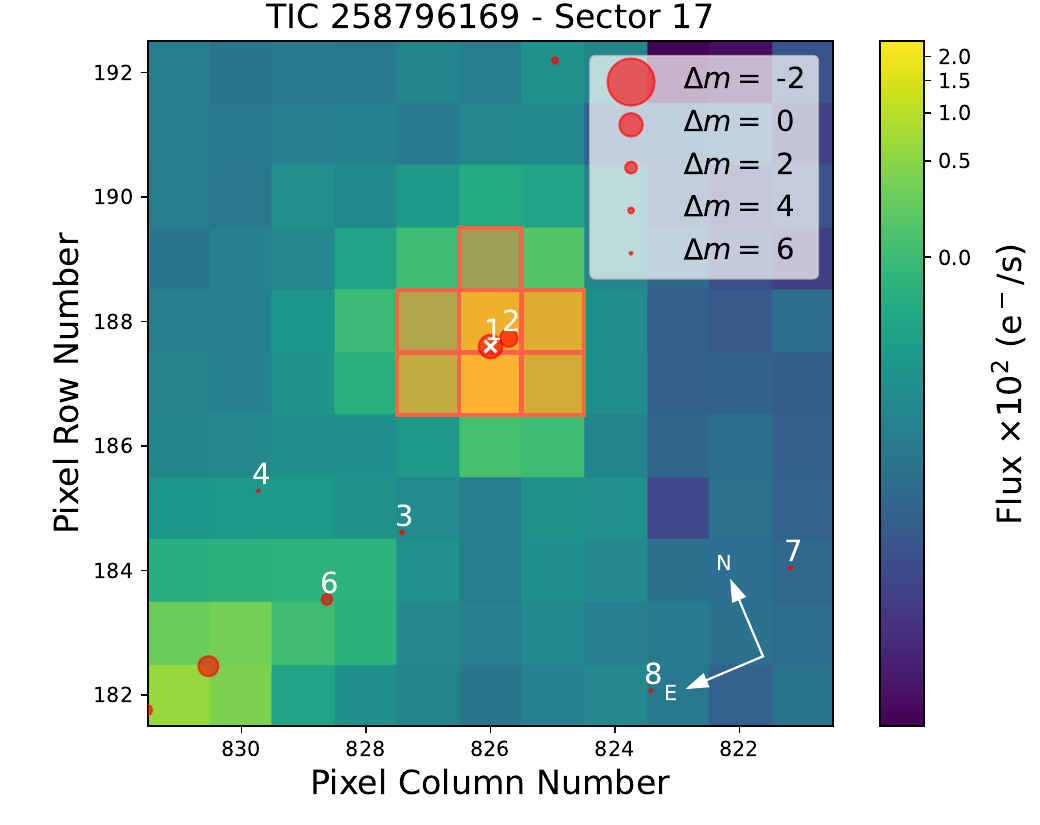}
      \caption{Target Pixel File (TPF) of TOI-4616 from TESS Sector 17.
The color scale shows the median pixel flux. The red squares indicate the optimal photometric aperture used to extract the light curve. Nearby sources are marked and scaled according to their TESS magnitude difference relative to the target ($\Delta m$). A neighbouring star falls within the photometric aperture and contributes a small amount of contaminating flux, which is accounted for in the analysis. The figure was produced using \texttt{TPFPLOTTER} \citep{aller2020planetary}.}
  \label{fig:TESStpf}
   \end{figure}

\section{Observations}
\label{sec:Observations}

\subsection{\textbf{TESS photometry}} 
\label{sec:tess}

\begin{figure*}
\centering
\includegraphics[width=\textwidth]{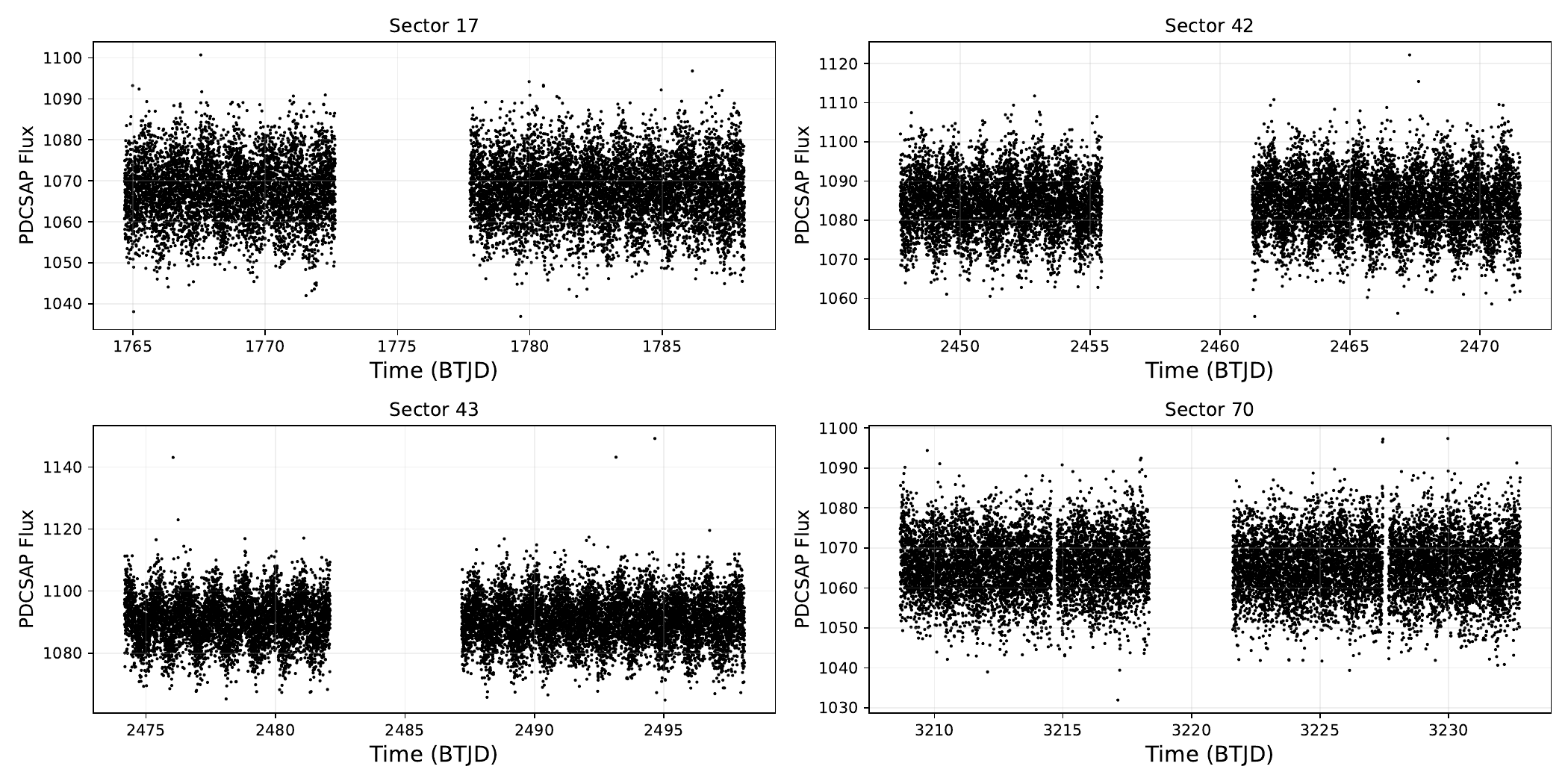}
\caption{TESS PDCSAP light curves of TOI-4616 from Sectors 17, 42, 43, and 70. 
Each panel shows the flux time series for one sector, illustrating the sector-to-sector variation in photometric scatter and long-term systematics. 
Data gaps reflect the two-orbit observing strategy of TESS.}
\label{fig:TESSlc}
\end{figure*}

TOI-4616 is included in the revised TESS Input Catalog and candidate target list \citep{stassun2018tess} and was observed at 2-minute cadence in Sectors 17, 42, 43, and 70. 
We extracted the PDCSAP light curves from all available sectors using the \texttt{LightKurve} package\footnote{\url{https://github.com/lightkurve}} \citep{2018ascl.soft12013L}. 
The sector light curves were stitched together, NaN values removed, and a $5\sigma$ clipping applied to reject outliers. 
The final dataset comprises more than 39000 photometric measurements and includes 68 transit events.

To reduce the computational cost of the joint transit analysis while preserving the information content of the data, we retained only measurements within $\pm10$ hours of each transit midpoint. 
This selection reduces the dataset to approximately 20000 points while maintaining full transit coverage and sufficient out-of-transit baseline.

\begin{figure*}
\centering
\includegraphics[width=\textwidth]{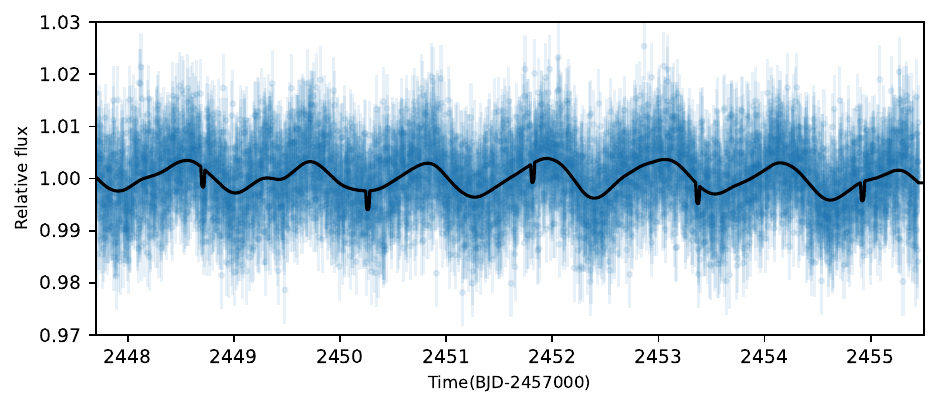}
\caption{Section of the TESS light curve illustrating the intrinsic variability of the host star. 
Blue points show the photometric measurements, while the black curve represents the Gaussian Process model used to capture correlated variability.}
\label{fig:TESSvariability}
\end{figure*}

The TESS light curve shows clear correlated variability, likely dominated by stellar rotation and instrumental systematics (Figure~\ref{fig:TESSvariability}). 
To account for these effects, we model the photometry using a Gaussian Process (GP) \ref{SBC:Noise} jointly with the transit model. 
All reported transit parameters are obtained while marginalizing over the GP hyperparameters in the global fit. 
For visualization purposes, we also display the GP-predicted variability and the corresponding detrended light curve.

\begin{figure}
\centering
\includegraphics[width=\columnwidth]{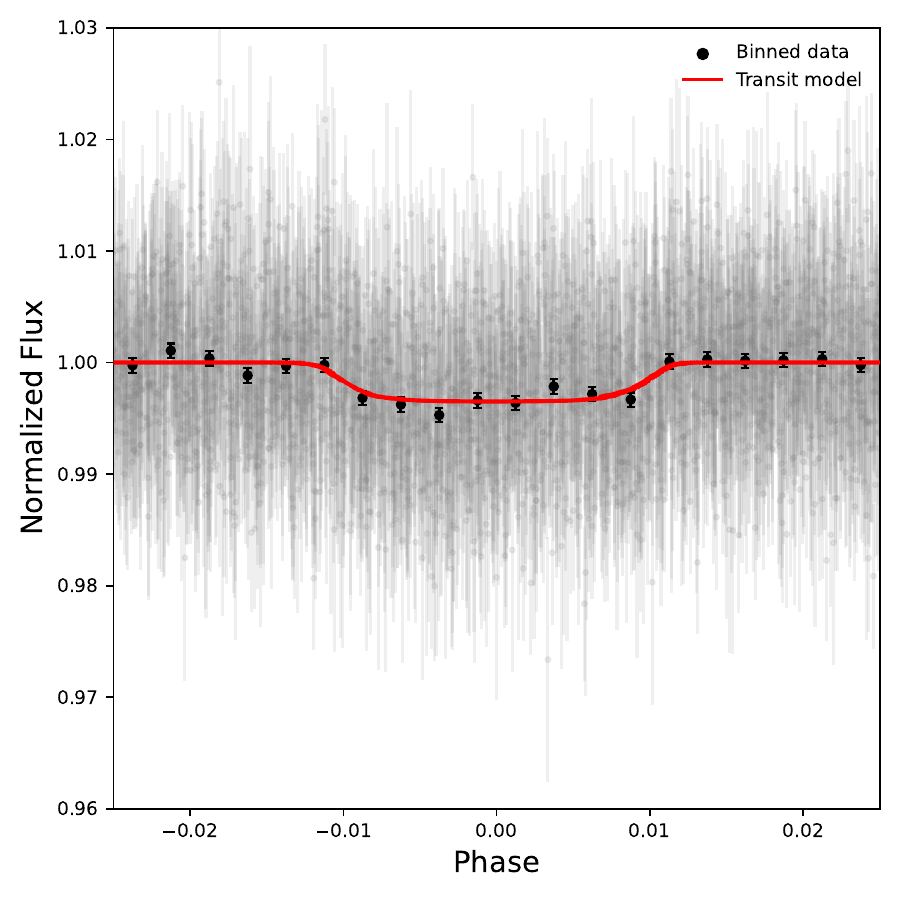}
\caption{Phase-folded TESS light curve after variability modeling. The grey
points show the GP-detrended photometry. The black points show the data
binned into 300 uniform bins in orbital phase from the full set of 54\,270
measurements. The red curve shows the transit model.}
\label{fig:TESSvGP_Detrended}\end{figure}

\subsection{\textbf{Ground based photometry}}
\label{subsec:Ground_based_phot}
    
\begin{table*}
\centering
\caption{Ground-based photometric follow-up observations for TOI-4616.}
\resizebox{\textwidth}{!}{%
\begin{tabular}{llccccccc}
\toprule
\textbf{Telescope (m)} & \textbf{Camera} & \textbf{Filter} & \textbf{Pixel scale (''/pix)} & \textbf{Aperture (pix)} & \textbf{Coverage} & \textbf{Date} & \textbf{Duration (min)} & \textbf{Exposures} \\
\midrule
SAINT-EX (1.0) & Andor ikon-L & I+Z& 0.35& 7.5 &Full transit & 2021 Nov 13 & 339 & 423 \\
SAINT-EX (1.0) & Andor ikon-L & I+Z& 0.35& 5.5 & Full transit & 2021 Nov 19 & 477 & 429 \\
SAINT-EX (1.0) & Andor ikon-L&I+Z & 0.35& 5.5& Full transit & 2021 Nov 27 & 421 & 525\\
SAINT-EX (1.0) & Andor ikon-L& I+Z& 0.35& 5.5& Full transit & 2021 Nov 30 & 444 & 569 \\
SAINT-EX (1.0) & Andor ikon-L & I+Z & 0.35& 5.5& Full transit & 2022 Jan 25 & 172 & 156 \\
 \hline
LCO (2.0)  & MuSCAT3 & i\'  & 0.27 & 5.1 & Full transit & 2022 Jan 15 & 564 & 188 \\
LCO (2.0)  & MuSCAT3 & r\'  & 0.27 & 5.1 & Full transit & 2022 Jan 15 & 480 & 160 \\
LCO (2.0)  & MuSCAT3 & g\' & 0.27 & 5.1 & Full transit & 2022 Jan 15 & 189& 63 \\
LCO (2.0) & MuSCAT3  & zs & 0.27 & 5.1  & Full transit& 2022 Jan 15 & 612 & 204 \\
\hline
TCS-MuSCAT2 (1.52)  & MuSCAT2 & i\'  & 0.44 & 4.0 & Full transit & 2021 Dec 05 & 237 & 456 \\
TCS-MuSCAT2 (1.52)  & MuSCAT2 & r\'  & 0.44 & 4.0 & Full transit & 2021 Dec 05 & 237 & 456 \\
TCS-MuSCAT2 (1.52)  & MuSCAT2 & g\' & 0.44 & 4.0 & Full transit & 2021 Dec 05 & 237 & 238 \\
TCS-MuSCAT2 (1.52) & MuSCAT2  & zs & 0.44 & 4.0  & Full transit& 2021 Dec 05 & 177 & 321 \\
LCOGT (1.0) & SINISTRO  & i\'  & 0.39 & 3.1  & Full transit & 2020 Mar 2 & 177 & 268 \\
\hline
SAINT-EX (1.0) & Andor ikon-L& i\'& 0.35& 8.0& Full transit& 2024 Oct 09 & 211 & 189\\
SAINT-EX (1.0) &Andor ikon-L & i\'& 0.35& 8.0& Full transit & 2024 Oct 20 & 238 & 250\\
\hline
SNO/Artemis (1.0) & Andor ikon-L & $z'$ & 0.35 & 5.0 & Full transit & 2025 Aug 03 &  200 & 310 \\

\bottomrule
\end{tabular}
\label{tab:toi4616_observations}
}

\end{table*}

\subsubsection{SAINT-EX photometry}

We observed TOI-4616 over seven nights with the 1-m SAINT-EX telescope at the San Pedro Mártir Observatory, Mexico, using the $I+Z$ on UTC 2021 November 13, 19, 27, 30, and UTC 2022 January 25 and $i^\prime$ bands on UTC 2024 October 09 and 20 with an exposure time of 30\,s. Observations were conducted using the \texttt{Astra} observatory control software \citep{astra} \footnote{https://github.com/ppp-one/astra}.

Raw science frames were reduced using the SAINT-EX module of the \texttt{PRINCE} pipeline \citep[Photometric Reduction and In-depth Nightly Curve Exploration;][]{demory2020super}, which performs bias, dark, and flat-field corrections. The world coordinate system (WCS)
solution was derived using \texttt{Astrometry.net} \citep{lang2010astrometry}. Reference stars were identified using \texttt{Photutils} \citep{bradley2020astropy}, and differential photometry was extracted via aperture photometry.

To correct for common-mode systematics (i.e., time-dependent trends affecting all stars in the field, such as transparency and airmass variations), we applied a weighted principal component analysis (WPCA; \citealt{delchambre2015weighted,Wells2021}) to the reference-star light curves. In this approach, each reference star is weighted by the inverse of its flux uncertainty, allowing the removal of shared trends while preserving astrophysical variability in the target light curve.

\begin{figure}
      \includegraphics[width=\columnwidth]{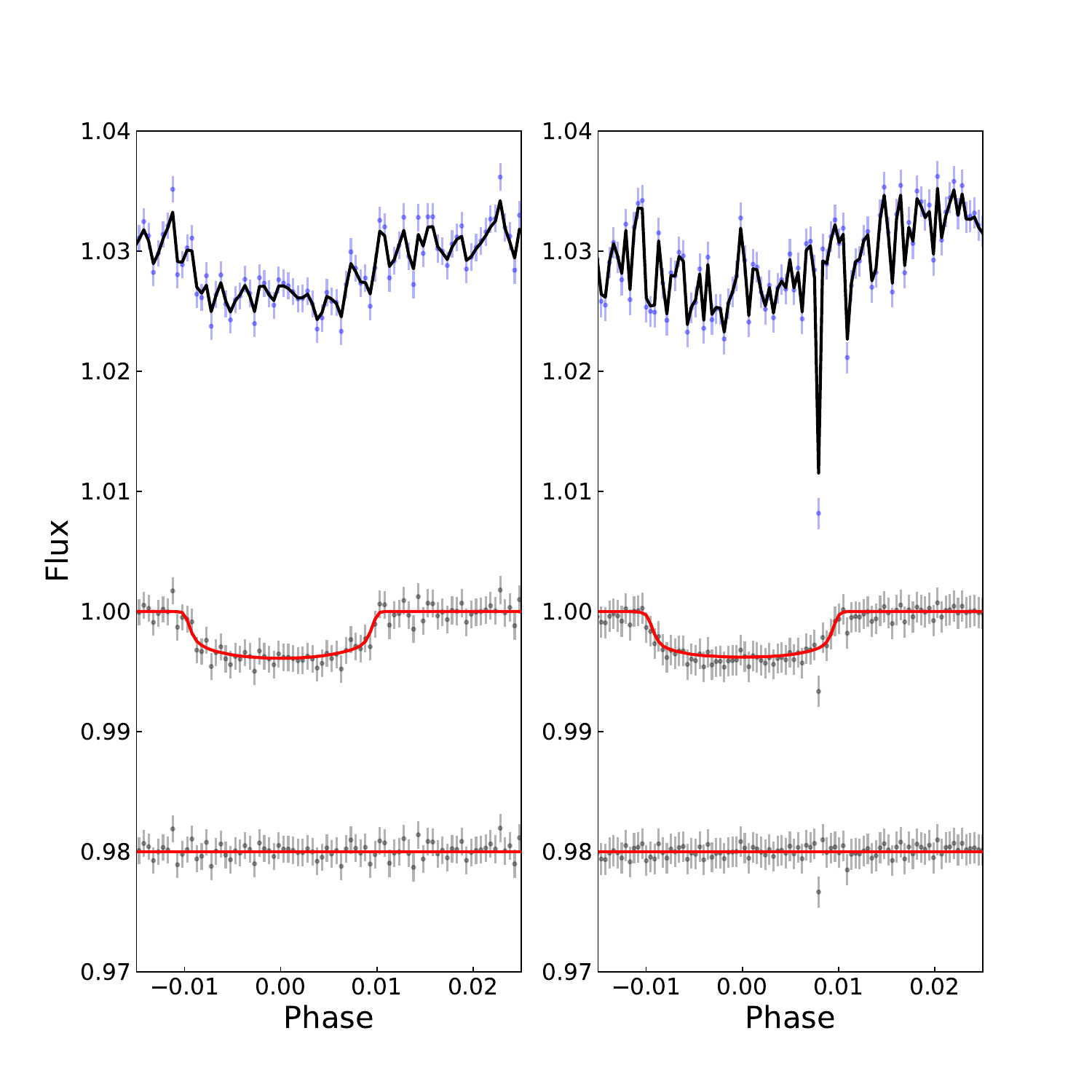}
      \caption{Recovery of the transit signal from highly variable SAINT-EX $i^\prime$-band photometry obtained on UTC 2024 October 09 (left) and UTC 2024 October 20 (right). In each panel, the top sub-panel shows the raw light curve (black points) and the GP model (blue curve) capturing correlated variability and systematics. The middle sub-panel shows the GP-detrended light curve with the best-fit transit model overplotted (red). The bottom sub-panel shows residuals after removing both GP and transit components. Despite strong time-correlated variability in the raw photometry, the joint GP+transit model robustly isolates the transit signal. The GP modeling was performed within the \texttt{juliet} framework.}
  \label{fig:SXIP}
\end{figure}

The resulting light curves exhibit significant correlated variability linked to stellar rotation and atmospheric systematics. We therefore modeled the data jointly with a Gaussian Process and transit model using the \texttt{juliet} framework, as illustrated in
Fig.~\ref{fig:SXIP}.

\subsubsection{\bf SPECULOOS-North/Artemis photometry}

SNO/Artemis \citep[Search for habitable Planets EClipsing ULtra-cOOl Stars,][]{Jehin2018Msngr,Delrez2018,Sebastian_2021AA,Burdanov2022} observed a full transit of TOI-4616\,b on UTC 2025 August 3 in the Sloan-$z'$ filter with an exposure time of 12s.
SNO/Artemis is equipped with a 2K$\times$2K Andor iKon-L camera with a pixel scale of $0.35\arcsec$, resulting in a field of view of $12\arcmin\times12\arcmin$.
Data reduction and photometric analysis were performed using the {\tt prose}\footnote{{\tt Prose:} \url{https://github.com/lgrcia/prose}} pipeline \citep{prose_2022}.

\subsubsection{\textbf{LCOGT photometry}}
  
We observed a full event window of TOI-4616 on UTC 2022 January 15 simultaneously in Sloan \(g', r', i'\), and Pan-STARRS $z$-short from the Las Cumbres Observatory Global Telescope \citep[LCOGT;][]{Brown:2013} \SI{2}{m} Faulkes Telescope North at Haleakala Observatory on Maui, Hawai'i. The telescope is equipped with the MuSCAT3 multi-band imager \citep{Narita:2020}. We also observed a full transit window on UTC 2021 December 09 from the LCOGT \SI{1}{m} network node at Teide Observatory on the island of Tenerife (TEID). We used the \texttt{TESS Transit Finder}, which is a customized version of the \texttt{Tapir} software package \citep{Jensen:2013}, to schedule our transit observations. All images were calibrated by the standard LCOGT \texttt{BANZAI} pipeline \citep{McCully_2018SPIE10707E} and differential photometric data were extracted using \texttt{AstroImageJ} \citep{Collins2017}. We used circular photometric apertures of \SI{5.1}{\arcsec} (MuSCAT3) and \SI{3.1}{\arcsec} (TEID) that excluded the flux from the nearest known neighbor in the Gaia DR3 catalog (Gaia DR3 2777125789169639168), which is \SI{6.8}{\arcsec} east of TOI-4616. The light curve data are available on the \texttt{EXOFOP-TESS} website\footnote{\label{fn:exofop}\url{https://exofop.ipac.caltech.edu/tess/target.php?id=258796169}}.

\subsection{MuSCAT2 photometry and transit analysis}

TOI-4616 was observed on UTC 2021 December 5, using the multi-band imager MuSCAT2 \citep{2019JATIS...5a5001N}, mounted on the 1.5\,m Telescopio Carlos S\'{a}nchez (TCS) at Teide Observatory, Spain. MuSCAT2 is equipped with four CCDs, enabling simultaneous imaging in the $g'$, $r'$, $i'$, and $z_s$ bands with minimal readout time. Each CCD features $1024 \times 1024$ pixels, providing a field of view of $7.4 \times 7.4$ arcmin$^2$. Observations were conducted at the telescope's nominal focus, with exposure times set to 30\,s for the $g'$, $r'$, and $i'$ bands, and 15\,s for the $z_s$ channel. The raw data were processed using the MuSCAT2 pipeline \citep{parviainen2019multicolour}, which performs dark and flat-field calibrations, aperture photometry, and transit model fitting, including corrections for instrumental systematics.

The MuSCAT2 dataset exhibits strongly correlated noise associated with atmospheric variability and site-dependent systematics, which prevents a stable recovery of the transit depth under a purely parametric model. To appropriately capture the structure of this noise, we employ a Gaussian Process with a quasi-periodic covariance kernel. The periodic term is tied across filters via the stellar rotation period inferred from the TESS data set Fig. \ref{fig:TESSvariability}, while the amplitude and length-scale hyperparameters are allowed to vary per filter to account for wavelength-dependent systematics. This approach yields a physically realistic characterization of the noise and suppresses spurious depth variations that would otherwise mimic chromaticity. We propagated the uncertainty on $d_{\rm rot}$ by sampling it with a Gaussian prior in the GP model.

Under this GP model, the MuSCAT2 transit depth converges to $d = 0.049 \pm 0.012$, which is fully consistent with the depth obtained from TESS and the higher-precision ground-based instruments \ref{Fig:MuSCAT2}. Consequently, although MuSCAT2 does not independently constrain chromaticity, it provides a valuable corroborative dataset that is compatible with the planetary interpretation when treated with an adequate noise model.

In the MuSCAT2 $i'$ observations, a single $>8\sigma$ outlier was identified and removed prior to the joint fit. Its inclusion introduced a spurious pre-ingress trend, whereas its removal yields a flat baseline without affecting the inferred transit depth.

\section{High resolution imaging}
\label{sec:High_res_imaging}

\subsection{Speckle imaging with NESSI at WIYN}
\label{subsec:Speckle_NESSI}
  
We observed TOI-4616 on UTC 2021 December 21 using the NN-EXPLORE Exoplanet Stellar Speckle Imager (NESSI) \citep{scott2018nn}, a speckle imager employed at the WIYN~3.5~m telescope on Kitt Peak.  NESSI was used to obtain speckle images in a filter at $\lambda_c = 832$~nm. The observation consisted of a set of 9000 exposures each with a 40~ms exposure time.  NESSI's field-of-view was limited by using $256\times256$ pixel sub-array readouts, resulting in a $4.6\arcsec\times4.6\arcsec$ field.  However, our speckle measurements were further confined to an outer radius of $1.2\arcsec$ from TOI-4616.  Speckle imaging of a point source standard star was taken in proximity to the observation of TOI-4616 and consisted of a set of 1000 exposures used to calibrate the intrinsic PSF.

These speckle data were reduced using the pipeline process described in \citet{howell2011speckle}.  Among the pipeline products is a reconstructed image of the field around TOI-4616.  We used this to measure contrast curves, setting detection limits on point sources close to the TOI-4616 host star.  No companion sources were detected within $1.2\arcsec$ of the star brighter than the contrast limits, which reached 3.9~magnitudes at $0.2\arcsec$ and 5~magnitudes at $1\arcsec$ from the star.

\begin{figure}
      \centering
      \includegraphics[width=\linewidth]{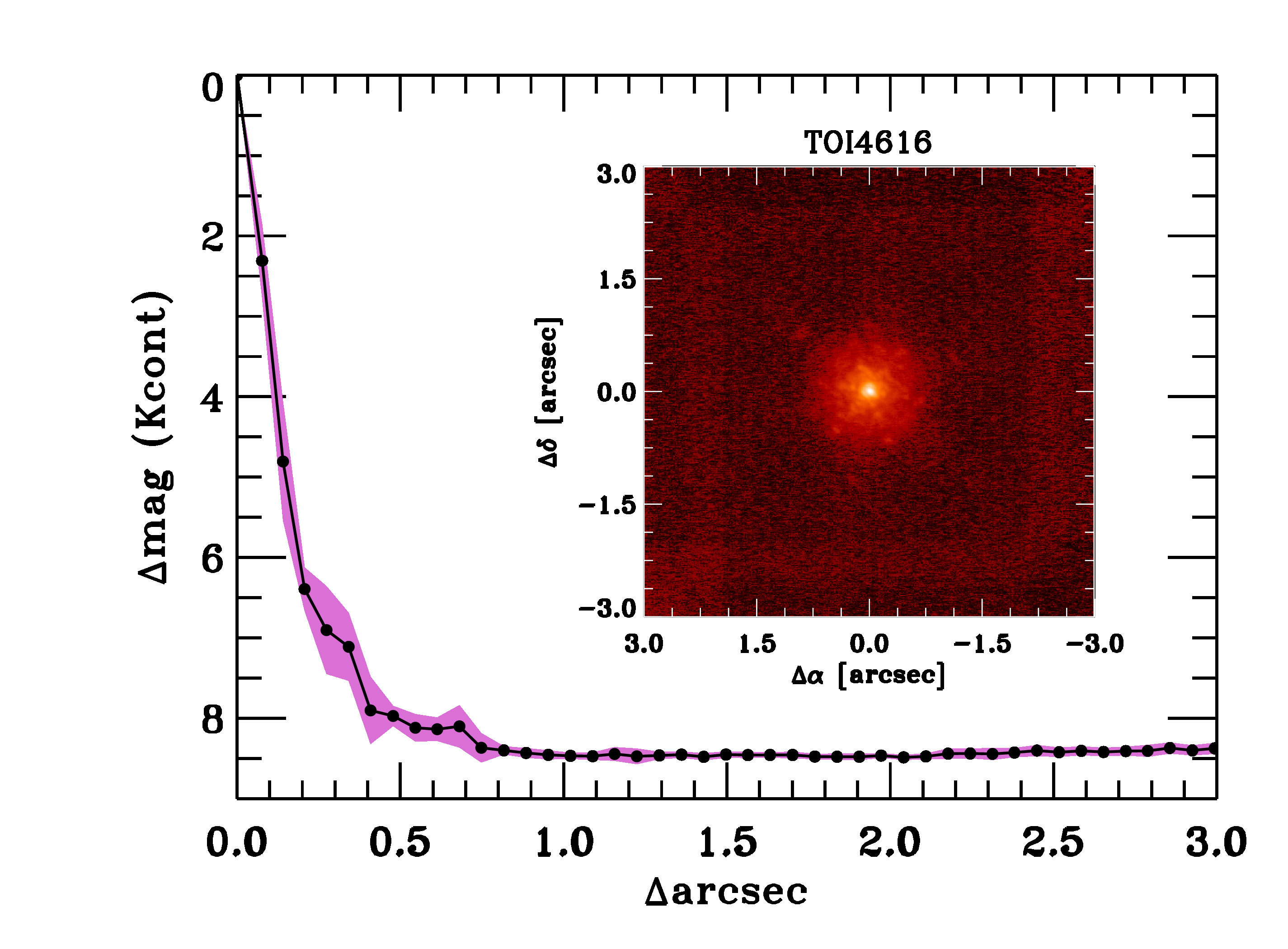}
      \caption{The contrast curve and the high-resolution image show the absence of any close companion.}
  \label{fig:Kcont}
\end{figure}

\section{Stellar Characterization}
\label{sec:stell_spect}

\subsection{Shane/Kast optical spectroscopy}

\begin{figure}
    \centering
    \includegraphics[width=\linewidth]{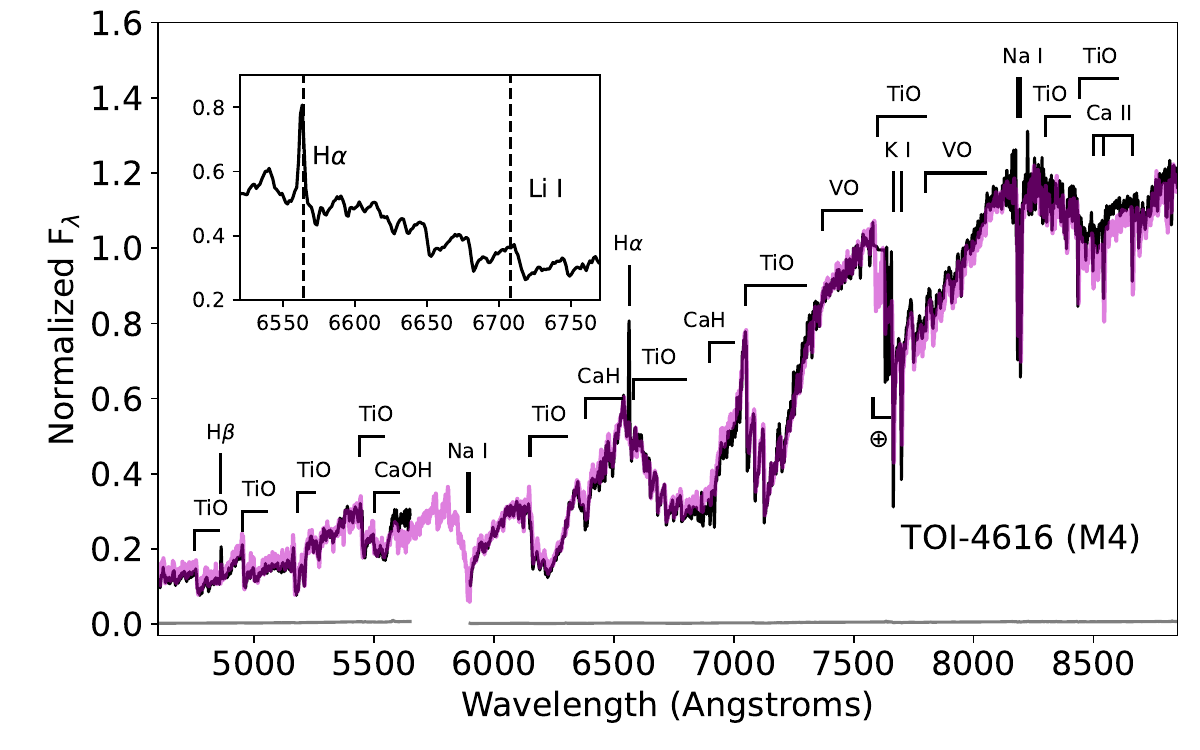}
\caption{Blue and red Shane/Kast optical spectrum of TOI-4616 (black line) compared to the best-fit M4 SDSS spectral template from \citet[magenta line]{2007AJ....133..531B}. Key spectral features are indicated, including regions of strong telluric absorption ($\oplus$). 
The inset box shows the 6520--6770~$\AA$ region encompassing the strong 6563~$\AA$ H$\alpha$ emission line and absent 6708~$\AA$ Li\,\textsc{i} absorption.}
    \label{fig:kast}
\end{figure}

We observed TOI-4616 with the Kast double spectrograph \citep{kastspectrograph} on the 3m Shane telescope at Lick Observatory on UTC 2021 November 27 in cloudy and windy conditions with 0$\farcs$9 seeing. We used the 1$\farcs$5 slit aligned to the parallactic angle to obtain blue and red optical spectra split at 5700~$\AA$ by the d57 dichroic, and dispersed by the 600/4310 grism and 600/7500 grating, resulting in spectral resolutions of $\lambda/\Delta\lambda$ $\approx1100$ and $\approx1500$, respectively. We obtained a single 1200\,s exposure in the blue channel and two 600\,s exposures in the red channel at an average airmass of 1.10. The G5\,V star HD 3266A ($V=8.3$) was observed afterward at a similar airmass for telluric absorption calibration, and the spectrophotometric calibrator Feige~110 \citep{1992PASP..104..533H,1994PASP..106..566H} was observed on the same night for flux calibration.  HeHgCd and HeNeArHg arc lamp exposures were used to wavelength calibrate our blue and red data, and flat-field lamp exposures were used to correct for pixel response variation. Data were reduced using the kastredux code\footnote{\url{https://github.com/aburgasser/kastredux}.} using standard settings. The resulting spectra have median signals-to-noise of 45 at 5425~$\AA$ and 167 at 7350~$\AA$.

The reduced spectrum is shown in Fig.\,\ref{fig:kast}, and is compared to the best-fit M4 dwarf spectral template from 
\citet{2007AJ....133..531B}. This optical classification is consistent with that reported by \citet{2019AJ....157...63K}, and confirmed using index-based methods described in 
\citet{1995AJ....110.1838R,1997AJ....113..806G,1999AJ....118.2466M,2003AJ....125.1598L}; and \citet{2007MNRAS.381.1067R} all of which yield M4 types.
We detect strong 6563~$\AA$ H$\alpha$ and 4861~$\AA$ H$\beta$
with equivalent widths of $-3.80\pm0.11$~$\AA$ and $-2.8\pm0.3$~$\AA$, respectively.
The former translates into a relative H$\alpha$ luminosity of $\log{\left(L_{{\rm H}\alpha}/L_{\rm bol}\right)} $=$ -3.86\pm0.06$ using the $\chi$ factor relation of \citet{2014ApJ...795..161D}, well into the saturated regime. 
The presence of strong H$\alpha$ emission indicates an activity age $\lesssim$4.5~Gyr \citep{2008AJ....135..785W} and possibly younger, although the absence of detectable 6708~$\AA$ Li\,\textsc{i} absorption rules out an age less than $\sim30$\,Myr. We measure a zeta metallicity index of $\zeta=0.939\pm0.003$  \citep{2013AJ....145..102L}, corresponding to a solar metallicity ([Fe/H] = $-0.07\pm0.20$; \citealt{2013AJ....145...52M}), consistent with prior measurements from APOGEE ([Fe/H] = $-0.071\pm0.009$; \citealt{2022AJ....163..152S}).

\subsection{IRTF/SpeX near-infrared spectroscopy}

\begin{figure}
    \centering
    \includegraphics[width=\linewidth]{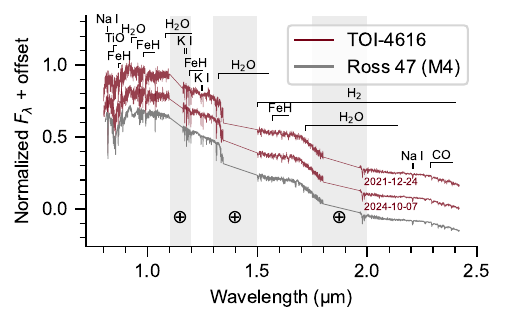}
    \caption{
    SpeX SXD spectra of TOI-4616 from 24 Dec 2021 (top) and 7 Oct 2024 (bottom).
    The target spectra (red) are shown alongside the M4V standard Ross 47 (grey), offset vertically for clarity.
    Prominent M-dwarf spectral features are labeled, and regions of strong telluric absorption are shaded.
    }
    \label{fig:spex}
\end{figure}

We observed TOI-4616 with the SpeX spectrograph \citep{Rayner2003} on the 3.2-m NASA Infrared Telescope Facility (IRTF) on UTC 2021 December 24  and UTC 2024 October 7.
Conditions were clear on both nights, with seeing of $1\farcs{4}$ and $1\farcs{2}$.
In each visit, we used the short-wavelength cross-dispersed (SXD) mode of SpeX with the  $0.3'' \times 15''$ slit aligned to the parallactic angle, which yields 0.80--2.42\,$\mu$m spectra at a resolving power of $R{\sim}2000$, and obtained six exposures while nodding in an ABBAAB pattern.
Integration times were 180\,s and 150\,s per exposure.
Each sequence was followed by SXD calibration frames and six short exposures of the A0\,V standard HD\,7215 (V=7.0).
Both datasets were reduced with Spextool v4.1 \citep{Cushing2004}.
The final spectra from 2021 and 2024, shown in Fig.\,\ref{fig:spex}, have median SNR per pixel values of 123 and 112, respectively, and an average of 2.5\,pixels per resolution element.

We compared the SpeX spectra of TOI-4616 to M dwarf standards in the IRTF Spectral Library \citep{Cushing2005, Rayner2009} using the SpeX Prism Library Analysis Toolkit \citep[SPLAT, ][]{splat}, finding the best match to the M4\,V standard Ross 47 on both nights.
We therefore adopt an infrared spectral type of M4.0$\pm$0.5.
Using SPLAT, we also measured the equivalent widths of the $K$-band Na\,\textsc{i} and Ca\,\textsc{i} features and the H2O--K2 index \citep{Rojas-Ayala2012}.
Applying the empirical relation of \citep{Mann2013}, we derive nearly identical iron abundance estimates from the 2021 and 2024 observations.
Taking the weighted mean of the two, we adopted a final metallicity of  $\mathrm{[Fe/H]} = -0.32 \pm 0.11$ dex, where the uncertainty reflects the systematic limit of the empirical relation.

\subsection{Comprehensive characterization of the TOI-4616 system}
\label{sec:Stell_charact}
To derive accurate stellar properties for TOI-4616, we combine photometric, spectroscopic, and astrometric constraints with empirical calibrations appropriate for mid-M dwarfs. In this section we present the derivation of the stellar mass, radius, effective temperature, luminosity, and metallicity, which form the basis for determining the planetary parameters.

\subsubsection{Empirical Relations}

We estimated the stellar parameters of TOI-4616 using the empirical relations of \citet{Mann2015, Mann2019}.
Based on the apparent $K_s$ magnitude of the star \citep{Skrutskie2006} and its Gaia DR3 parallax \citep{GaiaCollaboration2023}, we derive a mass of $0.1611 \pm 0.0037\,M_\odot$.
Incorporating the metallicity measured from the SpeX spectrum, we estimate a radius of $0.1969 \pm 0.0056\,R_\odot$.
These values yield a surface gravity of $\log g = 5.057 \pm 0.027$\,(cgs).
The Gaia $BP$--$RP$ color corresponds to an effective temperature of $3280 \pm 77$\,K.
Using the Stefan–Boltzmann Law, we calculate a bolometric luminosity of $0.00404 \pm 0.00044\,L_\odot$.

\subsubsection{SED Analysis}

\begin{table}
\centering
\caption{Basic information of TOI-4616.
Stellar density $\rho_{\star,\mathrm{SED}}$ is computed from the quoted
$M_\star$ and $R_\star$; the transit-inferred density
$\rho_{\star,\mathrm{tr}}$ is reported separately in
Table~\ref{tab: jointFitPriors}.}
\label{tab:toi530}

\resizebox{\columnwidth}{!}{

\begin{tabular}{lll}
\toprule
\textbf{Parameter} & \textbf{Value} &  \\
\midrule

\multicolumn{3}{l}{\textit{Main identifiers}} \\
TOI & 4616 & \\
TIC & 258796169 & \\
Gaia ID & 2777125789169639296 & \\

\multicolumn{3}{l}{\textit{Equatorial coordinates}} \\
RA (J2000) & 00:59:16.34 & \\
Dec (J2000) & +13:51:38.86 & \\

\multicolumn{3}{l}{\textit{Photometric properties}} \\
TESS (mag) & 12.8872 $\pm$ 0.0074 & TIC V8\textsuperscript{a} \\
Gaia $G$ (mag) & 14.1881 $\pm$ 0.0006 & Gaia DR3\textsuperscript{b} \\
Gaia BP (mag) & 15.814 $\pm$ 0.004 & Gaia DR3 \\
Gaia RP (mag) & 13.538 $\pm$ 0.002 & Gaia DR3 \\
B (mag) & 15.83 $\pm$ 0.031 & APASS \\
V (mag) & 14.753 $\pm$ 0.039 & APASS \\
J (mag) & 11.317 $\pm$ 0.029 & 2MASS \\
H (mag) & 10.713 $\pm$ 0.033 & 2MASS \\
K (mag) & 10.47 $\pm$ 0.02 & 2MASS \\
WISE 3.4$\mu$m (mag) & 10.235 $\pm$ 0.022 & WISE \\
WISE 4.6$\mu$m (mag) & 10.018 $\pm$ 0.019 & WISE \\
WISE 12$\mu$m (mag) & 9.862 $\pm$ 0.052 & WISE \\
WISE 22$\mu$m (mag) & 8.652 & WISE \\

\multicolumn{3}{l}{\textit{Astrometric properties}} \\
$\pi$ (mas) & 35.5096 $\pm$ 0.0302 & Gaia DR3 \\
$\mu_{\alpha*}$ (mas yr$^{-1}$) & 31.509 $\pm$ 0.036 & Gaia DR3 \\
$\mu_{\delta}$ (mas yr$^{-1}$) & 179.276 $\pm$ 0.025 & Gaia DR3 \\

\multicolumn{3}{l}{\textit{Derived parameters}} \\
Distance (pc) & 28.17 $\pm$ 0.02 & This work (Gaia DR3) \\
$M_\star$ ($M_\odot$) & 0.1881 $\pm$ 0.0094 & This work \\
$R_\star$ ($R_\odot$) & 0.1889 $\pm$ 0.0096 & This work \\
$\log g$ (cgs) & 4.70 $\pm$ 0.03 & This work \\
$\rho_\star$ ($\rho_\odot$) & 28.0 $\pm$ 4.5 & This work \\
SpT & M4 $\pm$ 0.5 & This work \\
$T_{\rm eff}$ (K) & 3150 $\pm$ 75 & This work \\
$F_{\rm bol}$ ($\mathrm{erg\,s^{-1}\,cm^{-2}}$) & $(1.28 \pm 0.05)\times10^{-10}$ & This work \\
$\mathrm{[Fe/H]}$ & $-0.32 \pm 0.11$ & SpeX (adopted) \\
 & $-0.25 \pm 0.25$ & SED \\
 & $-0.071 \pm 0.009$ & APOGEE \\
$P_{\rm rot}$ (days) & 1.2 & This work \\
Estimated age (Gyr) & 0.3 -- 0.8 & This work \\

\bottomrule
\end{tabular}
}

\footnotesize{
\textsuperscript{a}TIC V8, \textsuperscript{b}Gaia DR3 \citep{gaia2021a}.
}
\end{table}

\begin{figure}
    \centering
\includegraphics[width=0.9\linewidth,trim=70 60 50 50,clip]{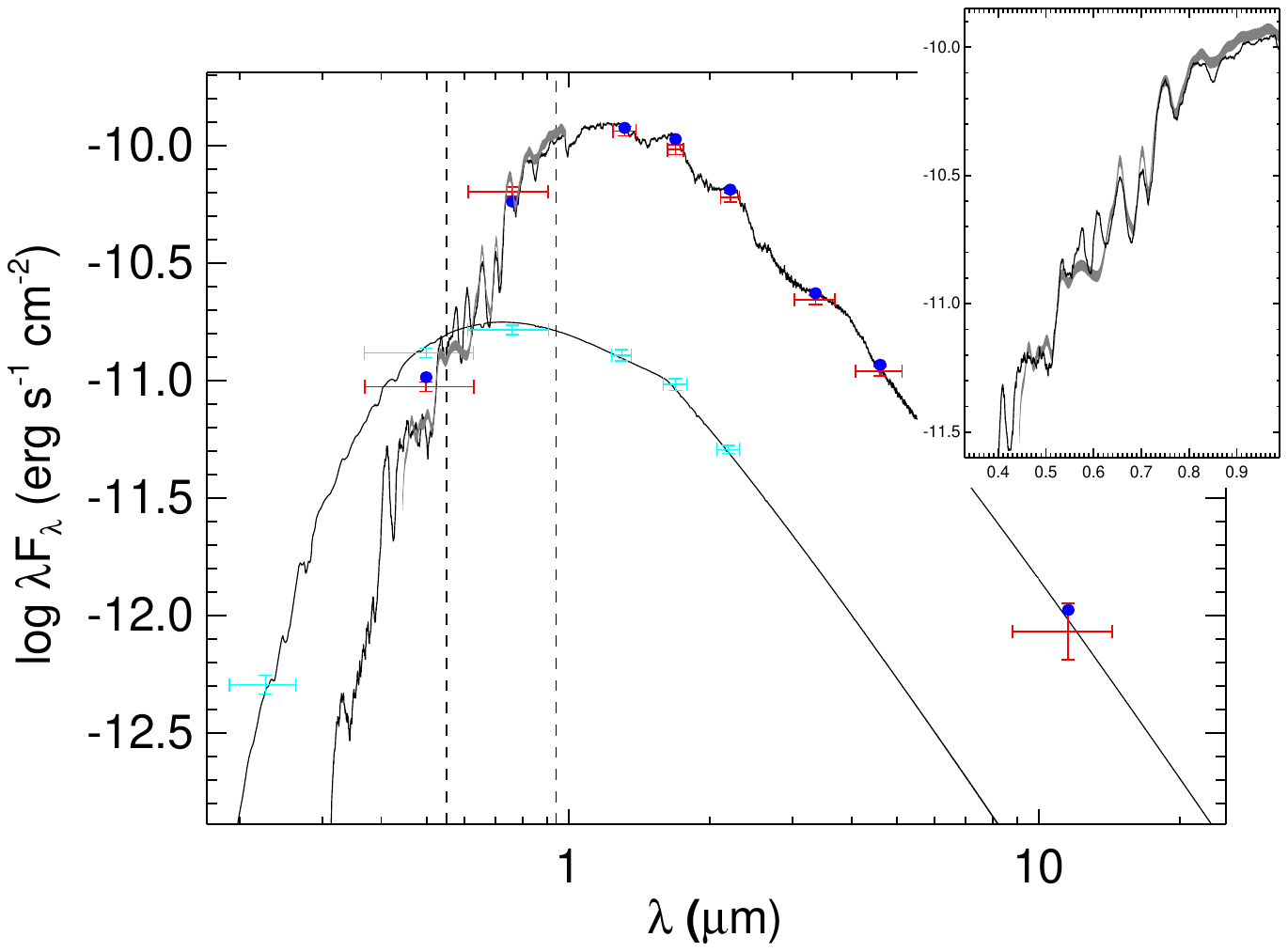}
\caption{Spectral energy distribution of TOI-4616. Red symbols represent the observed photometric measurements, where the horizontal bars represent the effective width of the passband. Blue symbols are the model fluxes from the best-fit PHOENIX atmosphere model (black). The {\it Gaia\/} spectrophotometry is represented by the grey swathe, also shown in greater detail in the inset plot. An unassociated bright star in the TESS photometric aperture is represented by the light blue data points and associated fitted SED model, with which we estimate the contaminating flux in the light curve of TOI-4616; see the text.\label{fig:sed}}
\end{figure}

As an independent determination of the basic stellar parameters, we performed an analysis of the broadband SED of the star together with the {\it Gaia\/} DR3 parallax \citep[with no systematic offset applied; see, e.g.,][]{StassunTorres:2021}, to determine an empirical measurement of the stellar radius, following the procedures described in \citet{Stassun:2016,Stassun:2017,StassunTorres:2018}. We pulled the the $JHK_S$ magnitudes from {\it 2MASS}, the W1--W3 magnitudes from {\it WISE}, the $G_{\rm BP} G_{\rm RP}$ magnitudes from {\it Gaia}, as well as the absolute flux-calibrated {\it Gaia\/} spectrophotometry. Together, the available photometry spans the full stellar SED over the wavelength range 0.4--10~$\mu$m (see Figure~\ref{fig:sed}).  

We performed a fit using PHOENIX stellar atmosphere models, with the free parameters being the effective temperature ($T_{\rm eff}$) and metallicity ([Fe/H]), as well as the extinction $A_V$, which we fixed at zero due to the proximity of the TOI-4616 system to Earth. While model uncertainties exist for cool stellar atmospheres at longer wavelengths, PHOENIX models provide an adequate representation of the broadband spectral energy distributions of mid-M dwarfs for the purpose of estimating the bolometric flux (e.g., \citep{Stassun:2016;Stassun:2017}). The resulting fit (Figure~\ref{fig:sed}) has a best-fit $T_{\rm eff} = 3150 \pm 75$~K, [Fe/H] = $-0.25 \pm 0.25$, with a reduced $\chi^2$ of 1.3. Integrating the (unreddened) model SED gives the bolometric flux at Earth, $F_{\rm bol} = 1.278 \pm 0.045 \times 10^{-10}$ erg~s$^{-1}$~cm$^{-2}$. Taking the $F_{\rm bol}$ and $T_{\rm eff}$ together with the {\it Gaia\/} parallax, gives the stellar radius, $R_\star = 0.1889 \pm 0.0096$~R$_\odot$. In addition, we can estimate the stellar mass from the empirical $M_K$ relations of \citet{Mann2019}, giving $M_\star = 0.1881 \pm 0.0094$~M$_\odot$. 

We identified an unassociated bright star in the TESS photometric aperture, therefore, we also fitted its SED to estimate its contribution of contaminating flux to the TESS light curve of TOI-4616 \citep{stassun2018tess, livingston2018tess, crossfield2016k2, mayo2018k2, collins2017exofastv2}. It is also represented in Figure~\ref{fig:sed}, with the following best-fit parameters: 
$T_{\rm eff} = 5250 \pm 100$~K,
$A_V = 0.24 \pm 0.02$, and
[Fe/H] $ = -2.0 \pm 0.5$, with a resulting
$R_\star = 5.8 \pm 1.0$~R$_\odot$. 

The stellar parameters derived from the empirical relations of \citet{Mann2015, Mann2019} and from the SED analysis are broadly consistent within the expected systematic uncertainties of empirical calibrations for mid-M dwarfs. In the remainder of this work we adopt the SED-based stellar parameters, which incorporate the measured bolometric flux and Gaia parallax and therefore provide a direct empirical constraint on the stellar radius.

Integrating its SED gives a ratio of its flux in the TESS bandpass to that of TOI-4616 of $0.391 \pm 0.017$.

We obtained multiple metallicity estimates for TOI-4616 from independent methods. 
The broadband SED fit provides only a weak constraint on [Fe/H] , as metallicity is partially degenerate with effective temperature and model assumptions in cool stellar atmospheres. 
The APOGEE pipeline yields a near-solar metallicity with very small formal uncertainty; however, the APOGEE pipeline uncertainties for mid-M dwarfs can be underestimated due to model and calibration systematics in this regime \citep{passegger2022metallicities}. 
From the SpeX spectrum, we derive a metallicity using empirical near-infrared M-dwarf relations, which rely on metallicity-sensitive spectral features and are specifically optimized for late-type stars \citep{lindgren2016metallicity}. 
Because these calibrations are tailored to M dwarfs and provide a physically motivated estimate, we adopt the SpeX-based metallicity as our fiducial value in the remainder of the analysis, while the APOGEE and SED values are treated as consistency checks. In the following sections, we therefore adopt the SpeX-based metallicity of [Fe/H] = -0.32 $\pm$ 0.11 as the fiducial stellar metallicity.

The M4 dwarf TOI-4616 exhibits strong H$\alpha$ emission 

(EW $ = -3.80 \pm 0.11$~\AA); 
\[ \log( L_{\mathrm{H}\alpha}/L_{\mathrm{bol}}) = -3.86 \pm 0.06\]
but no detectable 6708~\AA\ Li\,\textsc{i} absorption 
(see Sect.~\ref{sec:High_res_imaging}).
Combined with the rapid rotation of the star ($P_{\rm rot} \approx 1.2$\,d) 
and the absence of a close companion, these indicators suggest a relatively young system. 
We therefore estimate an age of roughly $0.3$--$0.8$\,Gyr, though we note that age determinations for mid-M dwarfs remain intrinsically uncertain.


\section{Modeling of the Transit System}
\label{sec:modeling}

\subsection{Transit Model and Implementation}

We modeled the transit light curves using the \texttt{BATMAN} package \citep{Kreidberg2015}, which implements the analytical transit formalism of \citet{mandel2002analytic}. Exploration of the posterior parameter space was performed using the affine-invariant Markov Chain Monte Carlo sampler \texttt{EMCEE} \citep{ForemanMackey2013}.

\subsection{Limb Darkening Treatment}
We adopted a quadratic limb-darkening law \citep{mandel2002analytic}.
To construct informative priors on the limb-darkening coefficients,
we used the LDTK toolkit \citep{Parviainen2015}, which interpolates PHOENIX stellar atmosphere models \citep{husser2013new} to compute filter integrated intensity profiles. The wavelength ranges of each instrument’s filter were supplied to LDTK together with the adopted stellar parameters of TOI-4616 ($T_{\rm eff}=3150$ K, $\log g=5.0$, [Fe/H] = $-0.32$). The resulting Gaussian priors on the coefficients $(u_1,u_2)$ were used in the
transit modeling.

Because TOI-4616 b has an orbital period of only 1.55 d, tidal circularization is expected to be efficient. Observational studies show that short-period planets typically exhibit eccentricities consistent with zero, particularly when the data do not constrain small deviations from circularity. We therefore fixed the eccentricity to zero in the transit modelling, as allowing it to float is not justified by the data and may introduce biases in derived system parameters \citep{hadden2014densities, eylen2019orbital,shporer2014atmospheric}.

\subsection{Noise Modeling and Gaussian Processes}
\label{SBC:Noise}

Each dataset was modeled jointly with a Gaussian Process (GP) noise component to account for correlated systematics and stellar variability. The GP was implemented using the SHOTerm kernel from the \texttt{celerite} package \citep{foreman2017fast}, which describes correlated variability as a stochastically driven damped harmonic oscillator.

The GP and transit model were fitted simultaneously. The GP component captures correlated noise and variability, while the transit parameters are inferred marginalizing over the GP hyperparameters.

\begin{figure*} \centering \includegraphics[width=\textwidth, height=0.4\textheight, keepaspectratio]{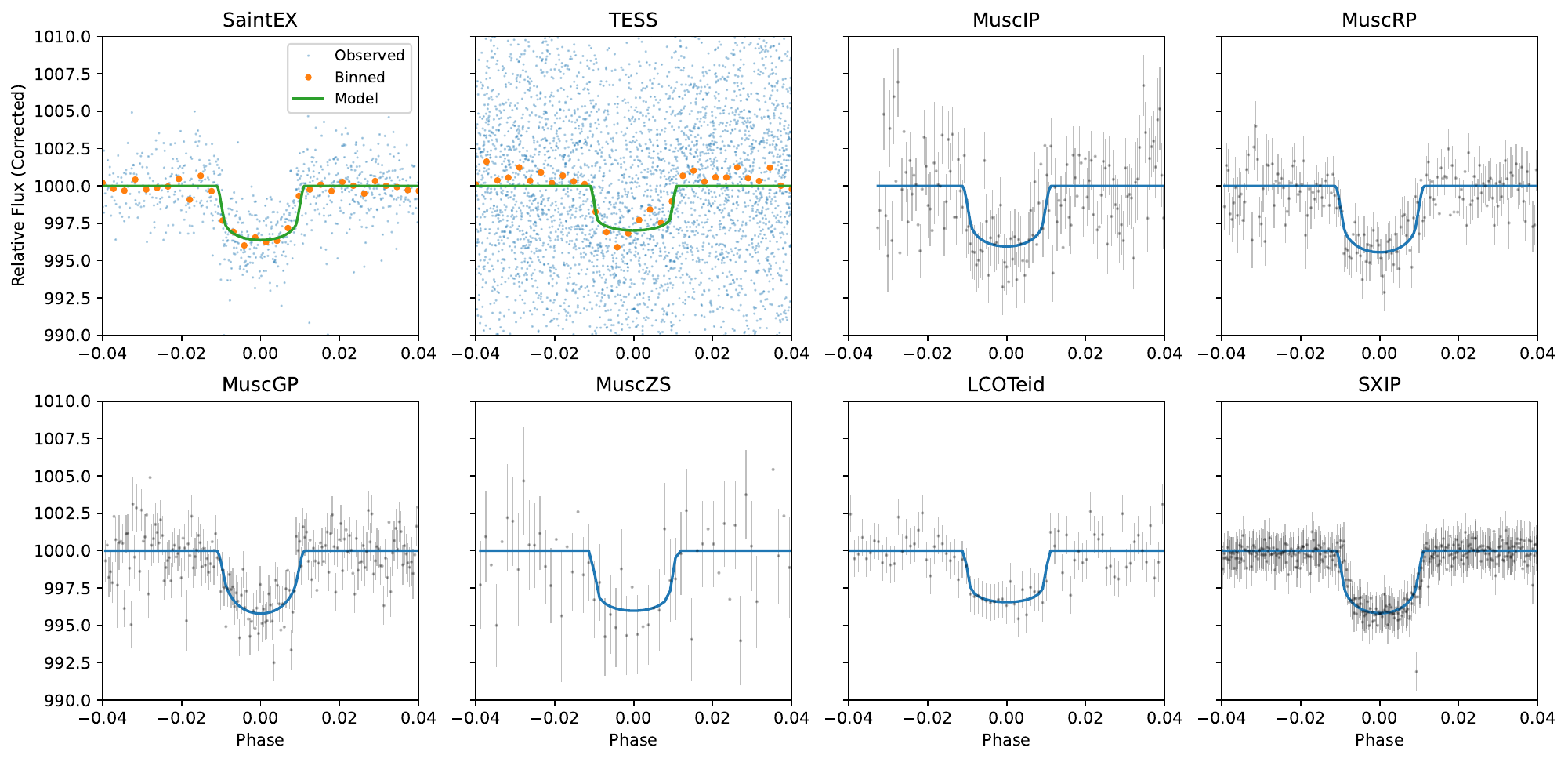} \caption{Phase-folded transit light curves and best-fit joint model for all available datasets: SAINT-EX, TESS, and the four MuSCAT3 filters ($g^\prime$, $r^\prime$, $i^\prime$, $z^\prime$). All panels are shown on identical phase and flux scales to allow direct comparison of transit depths. The solid line represents the median joint-fit model, while points show the observed data (with binning applied to the high-cadence SAINT-EX and TESS light curves for clarity).
The SXIP panel represents observations with the SAINT-EX telescope in i\' filter} \label{Fig: Modelfitstrans} \end{figure*}

\subsection{Multicolor Transit Fitting}

To investigate possible wavelength-dependent variations in transit depth, we allowed $R_{\rm p}$/$ R_\star$ to vary independently for each dataset. The impact parameter was constrained to physically meaningful values ($0 \le b \le 1 $+$ R_{\rm p} / R_\star$). For numerical stability, we parameterized $b$ through its absolute value to prevent unphysical negative solutions without altering the transit geometry.

Figure~\ref{Fig: Modelfitstrans} shows the phase-folded light curves and the median joint-fit model for all instruments. 
Small apparent differences in transit depth between datasets arise from photometric noise, filter-dependent systematics, and finite signal-to-noise ratios, and are consistent with the posterior uncertainties. 
We find no statistically significant wavelength-dependent variation in transit depth.

\begin{figure} \includegraphics[scale=0.47]{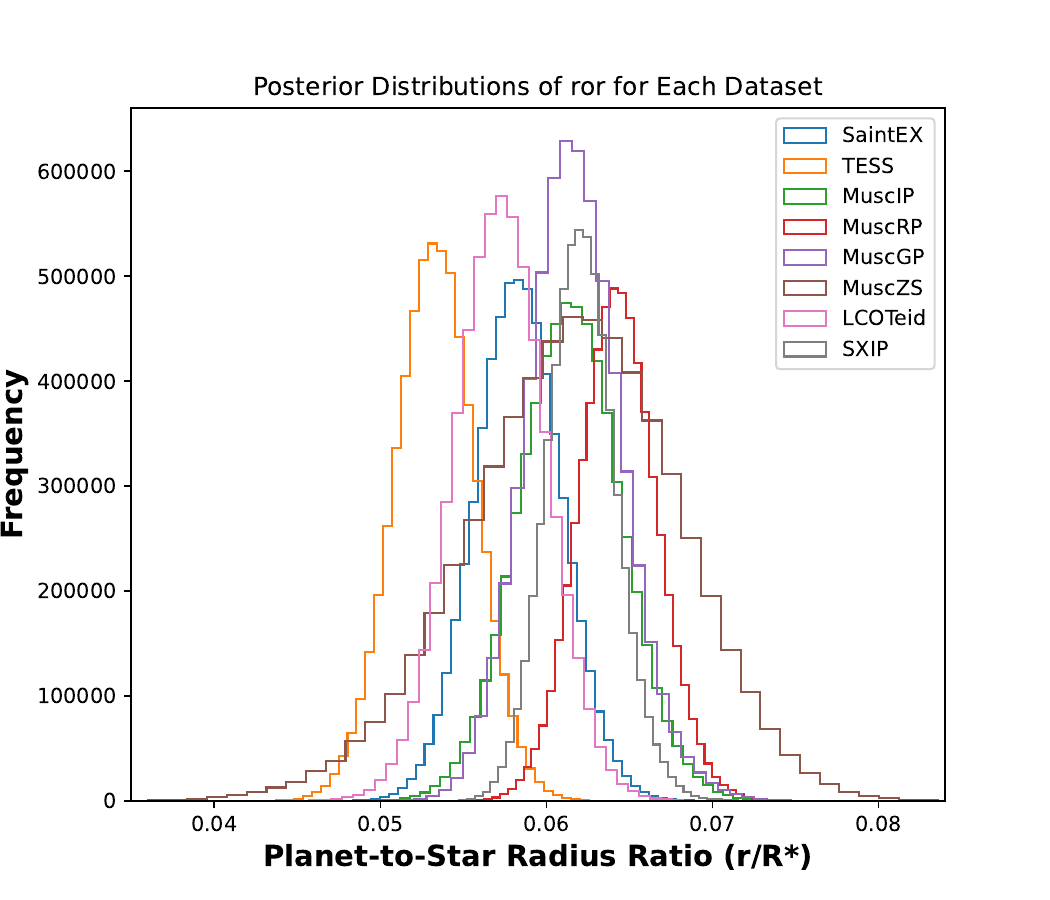} \caption{This figure shows the posterior distributions of the effective radius ratio for the six different filters used in our model. } \label{Fig:rorAll} \end{figure} 

\subsubsection{Radius ratio sampling} 

We extracted the posterior samples of the planet-to-star radius ratio, $R_{\rm p}/ R_\star$, from the joint multi-instrument fit shown in Fig.~\ref{Fig: Modelfitstrans}. 
The planet-to-star radius ratio distributions are shown in Fig.~\ref{Fig:rorAll}. 
In this fit, the transit depth was allowed to vary independently for each dataset in order to capture potential chromatic effects and instrument-dependent systematics. 

\begin{figure} \centering \includegraphics[scale=0.4]{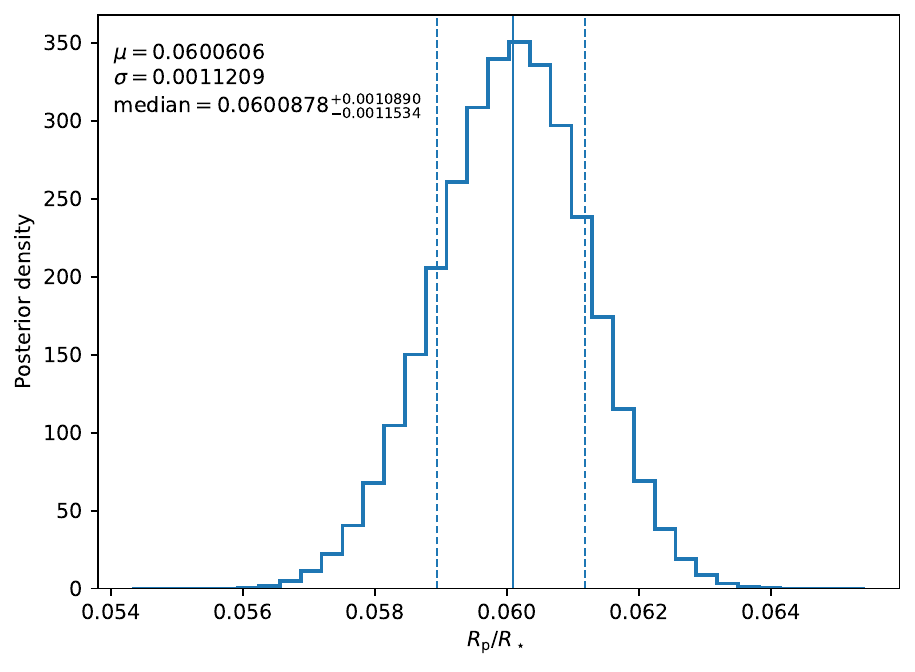} \caption{Posterior distribution of the planet-to-star radius ratio, $R_{\rm p}/R_\star$, obtained from the joint multi-instrument fit in which the transit depth was allowed to vary independently for each dataset (to capture potential chromatic effects). 
The distribution corresponds to the global posterior derived from the simultaneous fit to all datasets. 
The solid vertical line marks the posterior median, while the dashed lines indicate the 16th and 84th percentiles. 
The width of the distribution reflects the combination of statistical uncertainties and inter-instrument (and/or wavelength-dependent) depth variations.} \label{Fig: final_ror} \end{figure}

To obtain a representative global radius ratio, we combined the posterior samples from all eight datasets and constructed the aggregated posterior distribution shown in Fig.~\ref{Fig: final_ror}. 
Because the datasets were fitted simultaneously within a joint Bayesian framework, the resulting global posterior reflects the radius ratio consistent with all datasets and therefore provides a tighter constraint than any individual dataset alone. 
From this distribution, we derive a median value of:

\[R_{\rm p}/R_\star = 0.06009^{+0.00109}_{-0.00115}\]

where uncertainties correspond to the 16th and 84th percentiles.
Because the transit depth was allowed to vary independently for each dataset, the combined posterior should be interpreted as an instrument-marginalized estimate of $R_{\rm p}/R_\star$, which captures both statistical uncertainty and any residual inter-dataset systematics. We verified that the individual posteriors overlap within their credible intervals and that no statistically significant chromatic trend is present; the combined posterior is therefore used only as a representative global value.

\section{The search for additional planets in the system}
\label{sec:additional}

\begin{figure}
    \centering
    \includegraphics[width=\columnwidth]{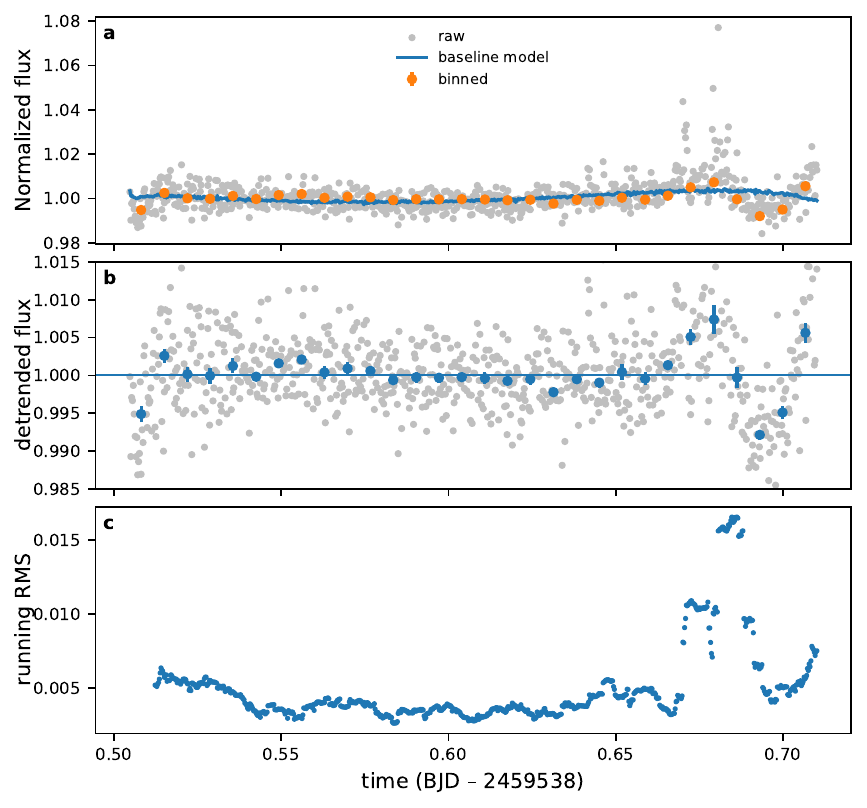}
    \caption{Ground-based photometry of TOI-4616 obtained with SNO/Ganymede on UTC 2021 November 20. (a) Raw differential flux and baseline model used for detrending. (b)Detrended light curve showing a shallow flux dip near the end of the observing sequence. (c) Running RMS of the detrended data, revealing a strong increase in correlated noise at the same time. The coincidence of the flux dip with the rise in photometric scatter suggests that the feature may be influenced by evolving systematics rather than a genuine transit.}
    \label{fig:lc_detrended}
\end{figure}

During our extended ground-based monitoring of TOI-4616, observations obtained with the SNO/Ganymede telescope on UTC 2021 November 20 revealed a transit-like feature that does not correspond to the ephemeris of TOI-4616\,b (Fig.~\ref{fig:lc_detrended}). The signal is visible in the raw differential photometry and remains present after detrending. The data were corrected for instrumental systematics using the PRINCE pipeline and independently reduced with AstroImageJ \citep{Collins2017}, yielding consistent light-curve features in the two independent reductions.

However, the event occurs near the end of the observing sequence, where correlations with airmass and other systematics increase and the photometric scatter rises. To quantify this behaviour, we computed a running RMS of the detrended residuals using a sliding window, which shows a clear increase in correlated noise toward the end of the sequence (Fig.~\ref{fig:lc_detrended}c). Independent reductions and tests with different sets of reference stars yield broadly consistent results, indicating that the feature is not obviously introduced by a particular reduction choice. Nevertheless, because it coincides with less favourable observing conditions—where increasing sky background and atmospheric variations can degrade photometric precision—its reliability remains uncertain. We therefore consider the feature suggestive but inconclusive. No other transit-like features were detected in our ground-based monitoring.

If interpreted as a planetary transit, the late-sequence flux dip in the SNO/Ganymede light curve has an approximate duration of $\sim$1--1.3\,h. For a planet orbiting TOI-4616 and assuming a circular orbit and near-central transit geometry, such a duration would correspond to an orbital period of order a few days ($\sim$2--5\,d), given the stellar density derived in Sect.~\ref{sec:Stell_charact}. However, the event occurs near the end of the observing sequence, where correlations with airmass increase and the photometric scatter rises. The coincidence of the dip with increased correlated noise, as indicated by the running RMS analysis (Fig.~\ref{fig:lc_detrended}), suggests that the feature is more plausibly associated with evolving systematics than with a genuine transiting signal.

Because the origin of this feature remains unclear, we analysed the TESS photometry to search for additional planets that could explain it. We used the \texttt{SHERLOCK} pipeline \citep{pozuelos2020,devora2024}, which is designed to identify planetary candidates that may be missed by standard detection pipelines such as SPOC or QLP \citep[see, e.g.,][]{geo2024,zuniga2025,seluck2025}. We performed two independent searches: one using 120\,s data from sectors 17, 42, 43, and 70, and another using 200\,s data from sectors 42 and 43. In both cases, we recovered the signal corresponding to TOI-4616.01, confirming its detectability. No additional planetary candidates were identified.

To further quantify our sensitivity to additional planets, we performed an injection--recovery analysis using the \texttt{MATRIX} code\footnote{\texttt{MATRIX} (\textbf{M}ulti-ph\textbf{A}se \textbf{T}ransits \textbf{R}ecovery from \textbf{I}njected e\textbf{X}oplanets) is available at \url{https://github.com/PlanetHunters/tkmatrix}} \citep{devora2022}. This framework explores the detection limits of the TESS data as a function of orbital period and planetary radius \citep[see, e.g.,][]{Wells2021,Schanche2021,pozuelos2023}. 

\begin{figure}
    \centering
    \includegraphics[width=\columnwidth]{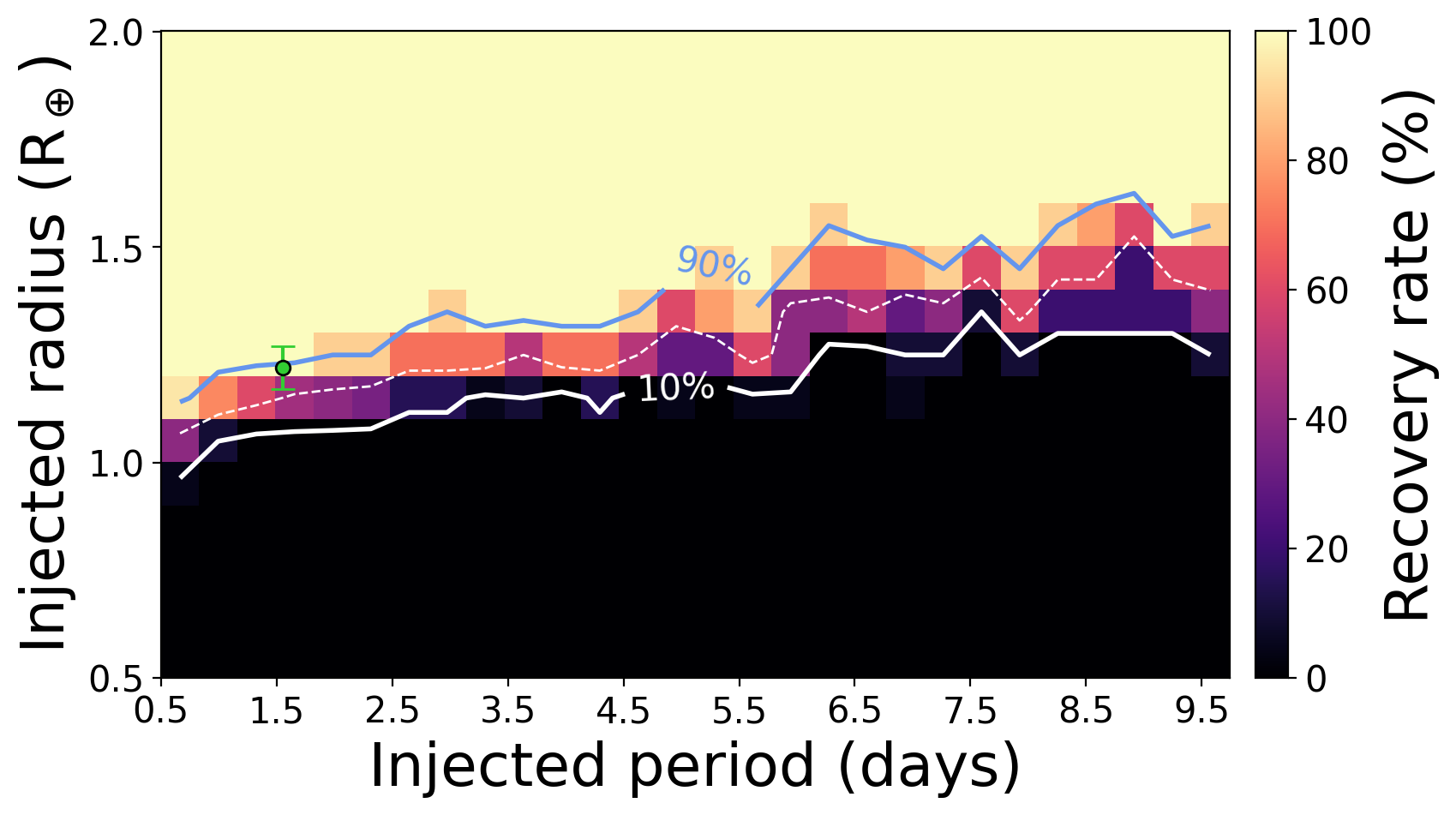}
    \caption{Results of the injection–recovery experiment conducted with \texttt{MATRIX}. The colour scale represents the recovery rate, with bright yellow indicating high recovery and dark purple/black indicating low recovery. The solid blue line marks the 90\% recovery contour, the dashed white line the 50\%, and the solid white line the 10\%. The green dot marks the nominal value for TOI-4616\,b.}
    \label{inj-rec}
\end{figure}

We constructed a grid of 30 orbital periods (0.5--10\,d), 30 planetary radii (0.5--3\,$R_{\oplus}$), and 10 transit epochs, yielding 9\,000 injected scenarios. For each case, a synthetic transit signal was injected into the original 120\,s TESS light curve and analysed after detrending with a bi-weight filter using a window size of 0.3\,d, which was found to be optimal during the \texttt{SHERLOCK} exploration. A planet was considered successfully recovered if the detected period and epoch deviated by less than 1\% and 1\,hour, respectively, from the injected values.

The results are shown in Fig.~\ref{inj-rec}. We identify a transition region where the recovery rate reaches 50\%; this threshold shifts from about 1\,$R_{\oplus}$ at an orbital period of 0.5\,d to roughly 1.3\,$R_{\oplus}$ at 10\,d. Planets smaller than 1\,$R_{\oplus}$ remain essentially undetectable across the explored period range, with recovery rates below 10\%. In contrast, planets larger than 1.5\,$R_{\oplus}$ are recovered with efficiencies above 80\%, allowing us to confidently rule out their presence.

These limits indicate that any additional transiting companion consistent with the ground-based feature would likely have a radius below $\sim1\,R_{\oplus}$ and therefore remain challenging to detect in the current TESS data.

\subsection{Transit Timing Variations (TTV)}

We measured individual mid-transit times for both the TESS and ground-based light curves while fixing the transit shape to the global solution. The resulting timing residuals are consistent with a linear ephemeris within uncertainties, and we find no evidence for significant TTVs in the current dataset. The timing analysis is shown in Appendix~\ref{app:TTV}.

\section{Validation of the planetary nature of the Object}
\label{sec:validation}

\subsection{Statistical validation}
We performed the statistical validation of TOI-4616\,b using the \texttt{TRICERATOPS}\footnote{\url{https://github.com/stevengiacalone/triceratops/tree/master}} package \citep{Triceratops_Giacalone_2021}, a Bayesian framework to evaluate the probability of the planet hypothesis. It does so by testing a variety of possible astrophysical origin scenarios for the transit event and calculating their relative probabilities based on the flux of nearby stars. The planetary hypothesis is then evaluated by the false positive probability (FPP), and the nearby false positive probability (NFPP): a planet is considered statistically validated if it has an FPP$< 0.015$ and an NFPP$<10^{-3}$. 

We performed an initial analysis using the four sectors of TESS data available (17,42,43,70). First, we removed the stellar variability signal using the \texttt{W{\={o}}tan}\footnote{\url{http://github.com/hippke/wotan}} package \citep{2019_Wotan_Hippke}. We selected a window size of 0.18 for the bi-weight filter to ensure that no transit feature is removed, and we computed an FPP of 0.823 and NFPP of 0.74. These probabilities are dominated by nearby eclipsing binary (NEB) and nearby transiting planet (NTP) scenarios around the closest neighboring star (TIC 258796170), which lies in the same pixels in TESS. We repeated the analysis with uncontaminated ground-based photometry, where a much smaller aperture was used to extract the target flux. Excluding the flux of the neighboring star removes the possibilities of NEB and NTP scenarios, effectively setting the NFPP to 0. By combining the contrast curve obtained from high-resolution imaging (see Section~\ref{subsec:Speckle_NESSI}) and the phase-folded light curve obtained in the Sloan-$i'$ filter (Section~\ref{subsec:Ground_based_phot}), the FPP was also reduced to 0.0135. These FPP and NFPP values are below the recommended thresholds (0.001 for NFPP and 0.015 for the FPP), and we can confidently classify TOI-4616 b as a validated planet. 

\subsection{Archival images}

\begin{figure*}
   \includegraphics[scale=0.31]{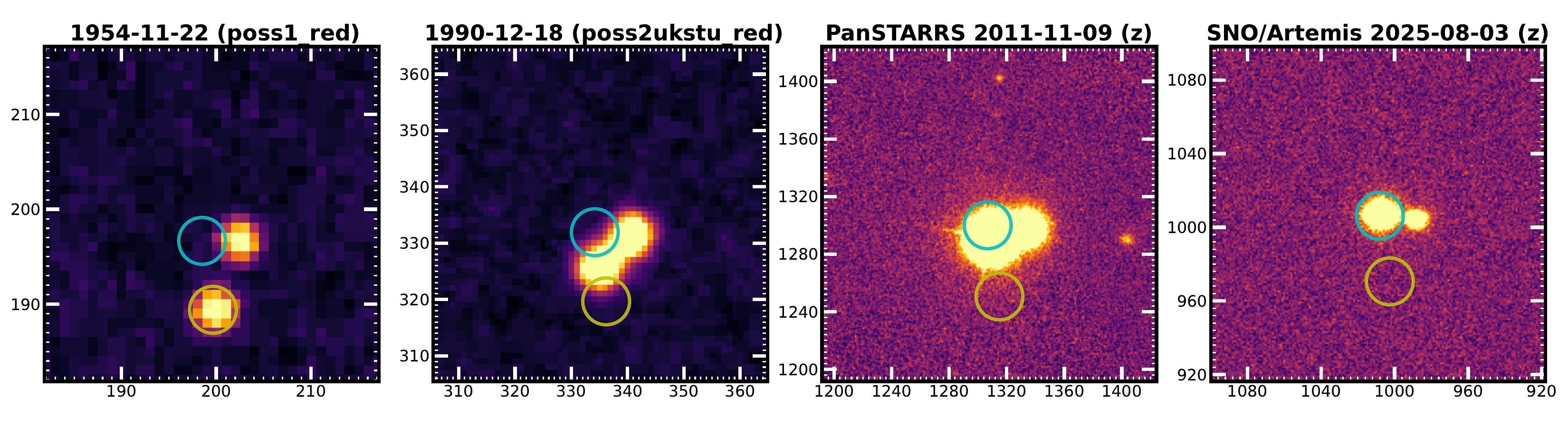}
   \caption{Archival images for TOI-4616. Yellow circles show the position of the star in the earliest image, and blue circles show the position of the star in the most recent images. The target shifted 12.9$^{\prime\prime}$ from 1954 to 2025. These images show that there is no background star blending within the current position of our target and that the nearby source is a background object.}
    \label{fig:archival_image}
\end{figure*}

We used the archival images for TOI-4616 to exclude any possible background stars that could be blended with our target at its current position. 
TOI-4616 has a proper motion of 182\,mas/yr. We used the data from POSS-I  in 1954 and POSS-II in 1990 in the red filter, and PanSTARRS in 2011 and SNO/Artemis in 2025 in the $z'$ filter, spanning a period of 71 years. The target has moved by $~\sim$12.9\farcs{4} from 1954 to 2025. Fortunately, no background star object is detected in the current position of TOI-4616 (see Fig.~\ref{fig:archival_image}).

\section{Discussion and Conclusions}
\label{sec:disc}

TOI-4616\,b resides in an extreme irradiation environment for an Earth-sized planet orbiting a mid-M dwarf. This makes it a particularly informative test case for models of atmospheric escape, interior composition, and volatile retention.

The combination of precise stellar parameters, consistent multi-band transit measurements, and the host star's brightness makes TOI-4616 a particularly valuable system for future atmospheric and dynamical studies. We therefore highlight this system as a benchmark terrestrial planet orbiting a mid-M dwarf, well-suited for comparative investigations of planetary structure and evolution in the strongly irradiated regime.

\begin{table*}
\centering
\caption{Joint-fit priors and best-fit parameters for TOI-4616.}
\label{tab: jointFitPriors}
\begin{threeparttable}
\renewcommand{\arraystretch}{0.70}
\setlength{\tabcolsep}{2pt}
\begin{tabular}{l c l c} 
\toprule
\textbf{Parameter} & \textbf{Prior} & \textbf{Best fit} & \textbf{Description} \\
\midrule
\multicolumn{4}{l}{\textit{Planetary parameters}} \\
$P_b$ (d) & $\mathcal{N} \left( 1.5538,\, 0.0001^{2} \right) $ & \num{1.55382(1)} & Orbital period \\
$T_{0,b}$ (BJD-2457000) & $\mathcal{N} \left(2684.880363943,\, 0.05^{2} \right)$ & \num{2684.88885(34)} & Mid-transit time \\
$b_{0}$ & $\mathcal{U} \left( 0,\, 1 \right)$ & \num{0.80(20)} & Impact parameter\\
$e_b$     & Fixed & 0 & Eccentricity \\
$\omega_b$ (deg) & Fixed & 90 & Arg. of periapsis \\
\midrule
\multicolumn{4}{l}{\textit{Stellar parameters}} \\
$\rho_\star$ ($\rho_\odot$) & $\mathcal{N} \left( 27.9,\, 2^{2} \right) $ & \num{30.655(02.700)} & Stellar density\\
\midrule
TESS Band & &  & TESS-specific parameters\\
$\mu$ & $\mathcal{N} \left( 0,\, 0.001^2 \right) $ & $0.0001 \pm 0.002 $ & mean flux\\
$u_1$ & $\mathcal{N} \left( 0.32,\, 0.15^2 \right) $ & \num{0.32(14)} & quadratic limb dark\\
$u_2$ & $\mathcal{N} \left( 0.26,\, 0.15^2 \right)$ & \num{0.29(14)} & linear limb dark\\
$d$ & $\mathcal{U} \left( 4.54\times 10^{-5}, 1.35\times 10^{-1} \right) $ & $0.0026 \pm 0.0005$ & Transit depth\\
\midrule
SAINT-EX I+z & &  & SAINT-EX-specific parameters\\
$\mu$ & $\mathcal{N} \left( 0,\, 0.001^2 \right)$ & $0.0000 \pm 0.0020$ & mean flux\\
$u_1$ & $\mathcal{N} \left( 0.29,\, 0.15^2 \right) $ & $0.33 \pm 0.14$  & quadratic limb dark\\
$u_2$ & $\mathcal{N} \left( 0.26,\, 0.15^2 \right) $ & $0.29 \pm 0.14$ & linear limb dark\\
$d$ & $\mathcal{U} \left( 1.2 \times 10^{-4}, 1.4 \times 10^{-1} \right) $ & $0.0031 \pm 0.0005$ & Transit depth\\
\midrule
MuSCAT3 Ip & &  & Ip-specific parameters\\
$\mu$ & $\mathcal{N} \left( 0,\, 0.001^2 \right) $ & $0.0000 \pm 0.0020$ & mean flux\\
$u_1$ & $\mathcal{N} \left( 0.32,\, 0.15^2 \right) $ & $0.37 \pm 0.14$ & quadratic limb dark\\
$u_2$ & $\mathcal{N} \left( 0.27,\, 0.15^2 \right) $ & $0.31 \pm 0.14$ & linear limb dark\\
$d$ & $\mathcal{U} \left(3.35 \times 10^{-4}, 5,00 \times 10^{-2} \right) $ & $0.0033 \pm 0.0007$ & Transit depth\\

\midrule
MuSCAT3 Rp & &  & Rp-specific parameters\\
$\mu$ & $\mathcal{N} \left( 0,\, 0.001^2 \right) $ & $0.0000 \pm 0.0020$ & mean flux\\
$u_1$ & $\mathcal{N} \left( 0.48,\, 0.15^2 \right) $ & $0.44 \pm 0.12$ & quadratic limb dark\\
$u_2$ & $\mathcal{N} \left( 0.24,\, 0.15^2 \right) $ & $0.26 \pm 0.13$ & linear limb dark\\
$d$ & $\mathcal{U} \left( 1.2 \times 10^{-4}, 1.4 \times 10^{-1} \right) $ & $0.0036 \pm 0.0004$ & Transit depth\\

\midrule
MuSCAT3 Gp & &  & Gp-specific parameters\\
$\mu$ & $\mathcal{N} \left( 0,\, 0.001^2 \right) $ & $0.0000 \pm 0.0020$ & mean flux\\
$u_1$ & $\mathcal{N} \left( 0.61,\, 0.15^2 \right) $ & $0.63^{+0.13}_{-0.12}$ & quadratic limb dark\\
$u_2$ & $\mathcal{N} \left( 0.29, 0.15^2\right) $ & $0.31 \pm 0.13$ & linear limb dark\\
$d$ & $\mathcal{U} \left( 1.2 \times 10^{-4}, 5.0 \times 10^{-2} \right) $ & $0.0031 \pm 0.0004$ & Transit depth\\

\midrule
MuSCAT3 z-short & &  & Z-Short-specific parameters\\
$\mu$ & $\mathcal{N} \left( 0,\, 0.001^2 \right) $ & $0.0000 \pm 0.002$ & mean flux\\
$u_1$ & $\mathcal{N} \left( 0.27, 0.15^2 \right) $ & $0.28 \pm 0.14$ & quadratic limb dark\\
$u_2$ & $\mathcal{N} \left( 0.26, 0.15^2 \right) $ & $0.26 \pm 0.15$ & linear limb dark\\
$d$ & $\mathcal{U} \left( 1.2 \times 10^{-4}, 5.0 \times 10^{-2}  \right) $ & $0.0034^{+0.0011}_{-0.0016}$ & Transit depth\\
\midrule
LCO-Teid ip & &  & Ip-specific parameters\\
mean & $\mathcal{N} \left( 0,\, 0.001^2 \right) $ & $0.0000 \pm 0.0020$ & mean flux\\
$u_1$ & $\mathcal{N} \left( 0.32,\, 0.15^2 \right) $ & $0.27 \pm 0.14$ & quadratic limb dark\\
$u_2$ & $\mathcal{N} \left( 0.27,\, 0.15^2\right) $ & $0.23 \pm 0.14$ & linear limb dark\\
$d$ & $\mathcal{U} \left( 1.2 \times 10^{-4}, 5.0 \times 10^{-2}  \right) $ & $0.0030 \pm 0.0006$ & Transit depth\\
\midrule
SAINT-EX ip & &  & ip-specific parameters\\
mean & $\mathcal{N} \left( 0,\, 0.001^2\right) $ & $0.0000 \pm 0.0020$ & mean flux\\
$u_1$ & $\mathcal{N} \left( 0.32,\, 0.15^2\right) $ & $0.42 \pm 0.12$ & quadratic limb dark\\
$u_2$ & $\mathcal{N} \left( 0.27,\, 0.15^2 \right) $ & $0.35 \pm 0.13 $ & linear limb dark\\
$d$ & $\mathcal{U} \left( 4.5 \times 10^{-5}, 5.0 \times 10^{-2}  \right) $ & $0.0034 \pm 0.0003$ & Transit depth\\

\bottomrule
\end{tabular}
\begin{tablenotes}
\footnotesize
\item \textbf{Note.} $\mathcal{U} \left(a, b \right) $: uniform prior from $a$ to $b$; $\mathcal{N} \left( \mu, \sigma^{2}\right) $: normal prior; quadratic limb darkening priors were calculated using the \texttt{LDTK} package \citep{Parviainen2015}. We took t0 at the midpoint of each dataset to remove the strong covariance between t0 and the period \citep{Eastman2013,Agol2005}.
\end{tablenotes}
\end{threeparttable}
\end{table*}

\begin{figure*}
   \centering
   \includegraphics[width=\textwidth,
                   height=0.8\textheight, keepaspectratio]{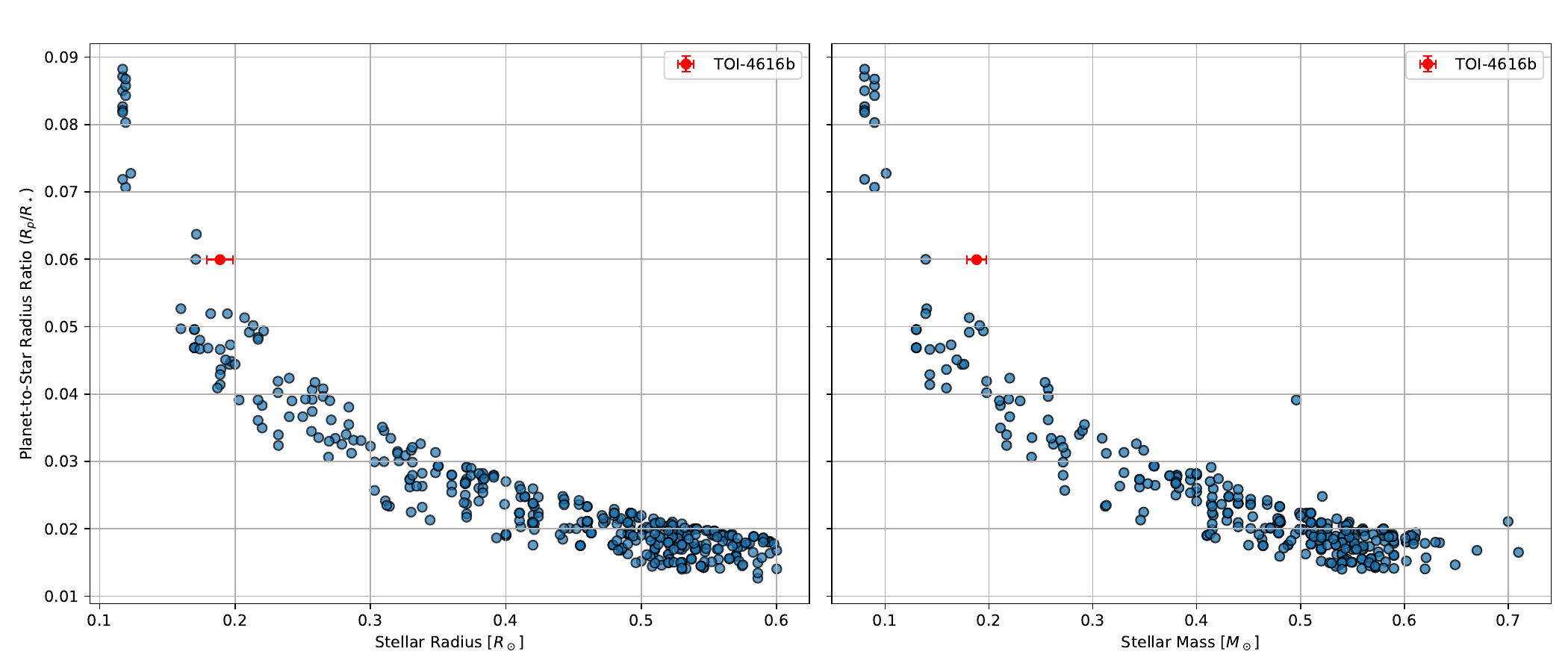}
      \caption{This shows the population of Earth-sized planets around M-dwarfs. In the right panel, we show the population in relation to stellar mass, and on the right, the population related to stellar radius. 
      The red dot in each panel shows the position of TOI-4616\,b relative to other planets. The data for the planetary and stellar populations are from the NASA EXOPLANET ARCHIVE accessed on UTC 2025 March 6 \citep{christiansen2022five}.
              }
         \label{Fig: planetPopulation}
\end{figure*}

\subsection{Mass-radius relation and escape velocity}
\label{sec:concl}

With a J-band magnitude of $J = 11.317 \pm 0.029$, TOI-4616 is accessible to high-precision near-infrared radial-velocity spectrographs optimized for mid-M dwarfs \citep{donati2018spirou}. Adopting a rocky mass–radius relation \citep{chen2017probabilistic}, we estimate a planetary mass of $M_p \approx 1.5$–$3.0\,M_\oplus$, implying a radial-velocity semi-amplitude of $K \approx 2.5$–$4.0\,\mathrm{m\,s^{-1}}$ \citep{suarez2017characterization}.

High-precision near-infrared spectrographs such as CARMENES \citep{schofer2021activity}, SPIRou \citep{karnavas2024derivation}, IRD, and NIRPS \citep{artigau2024nirps} are designed to achieve meter-per-second precision on mid-M dwarfs of similar brightness, suggesting that a mass determination may be feasible with an intensive monitoring campaign.

We performed Monte Carlo simulations to assess detectability. Assuming a semi-amplitude of $4.0\,\mathrm{m\,s^{-1}}$, per-epoch precision of $3.5\,\mathrm{m\,s^{-1}}$, and uniformly distributed observations over a 20-day baseline, we recover the signal at $\sim3.7\sigma$ with 20 epochs and $\sim5.7\sigma$ with 50 epochs. These estimates represent optimistic limits, as they do not include additional stellar variability.

The star’s rapid rotation ($P_{\rm rot} \approx 1.2$\,d inferred from the quasi-periodic timescale of the Gaussian-process model fitted to the TESS light curve) likely implies significant magnetic activity and RV jitter at the several $\mathrm{m\,s^{-1}}$ level on timescales comparable to the orbital period (1.55\,d). Disentangling planetary and activity signals will therefore require a substantial number of observations combined with activity diagnostics and joint modeling. We conclude that a precise RV mass measurement is challenging but potentially achievable with current facilities at the cost of significant telescope time.

\begin{figure}
   \centering
   \includegraphics[width=0.5\textwidth, height=0.4\textheight, keepaspectratio]{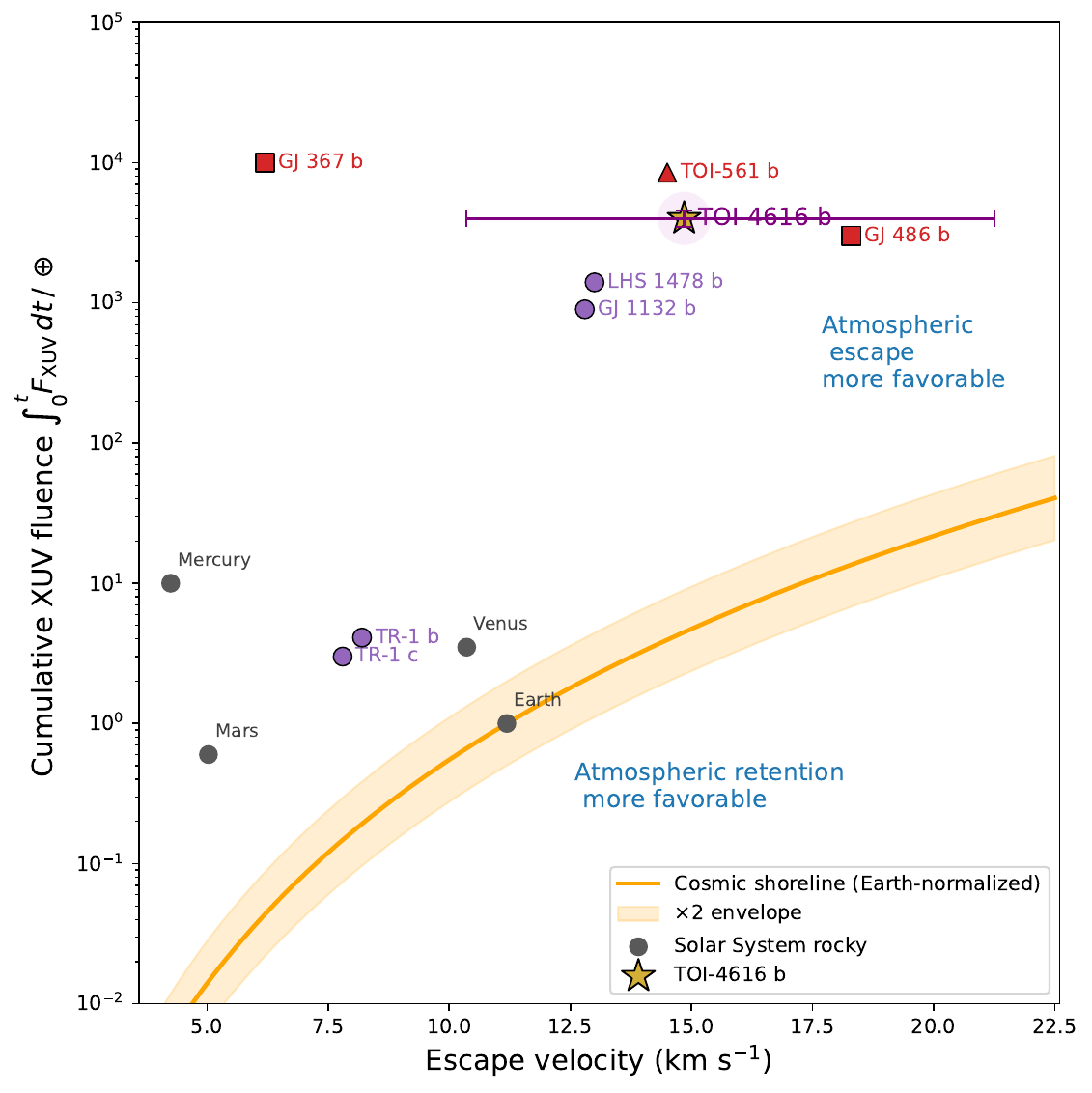}
      \caption{Location of TOI-4616 b in the cumulative XUV fluence–escape velocity plane, shown in the context of the empirical cosmic shoreline. The orange curve and shaded region indicate the Solar-System-derived cosmic shoreline relation \citep{zahnle2017cosmic} and its uncertainty envelope, separating regimes where atmospheres are expected to be retained (below) or efficiently eroded (above). Solar System rocky planets are shown for reference. TOI-4616 b (purple star) lies well above the shoreline, in a region where atmospheric survival is nominally disfavoured, at cumulative XUV fluence comparable to or exceeding that of TOI-561 b. The location of TOI-561 b—despite recent JWST evidence for a substantial atmosphere—highlights that the cosmic shoreline is not a strict boundary and motivates TOI-4616 b as a stringent test case for atmospheric escape and survival under extreme irradiation.}
         \label{Fig: shoreline}
\end{figure}

\subsection{Placement in the cosmic shoreline}

TOI-4616\,b lies in the high-irradiation, high-escape-velocity regime of the empirical cosmic shoreline \citep{ji2025cosmic,meni2025empirical}. Using the posterior distribution of the planetary radius and a rocky mass–radius relation, we estimate the planetary mass and compute the escape velocity as 
\[v_{\rm esc}=\sqrt{2GM_p/R_p}\], yielding 
\[v_{\rm esc}=14.9^{+6.4}_{-4.5} km s^{-1}\] 
Accounting for both bolometric instellation and stellar activity, we estimate the historical XUV irradiation using empirical activity–XUV scaling relations, obtaining 
\[J_{\rm XUV}/J_\oplus=(4.0\pm0.5)\times10^3.\]

with activity level
\[
\log\left(\frac{L_{\mathrm{H}\alpha}}{L_{\mathrm{bol}}}\right)
= -3.86 \pm 0.06 .
\]

A schematic comparison of equilibrium temperature and irradiation regimes for short-period rocky planets is shown in Appendix~D (Fig.~\ref{Fig: JWST_pers}).

The planet therefore, lies in a region where primordial H/He envelopes are expected to be efficiently stripped, while compact secondary atmospheres may survive (like GJ~486\,b, see Fig.\ref{Fig: shoreline}) \citep{misener2021cool}. Any present atmosphere is thus likely to be high-mean-molecular-weight (e.g.\ CO$_2$, N$_2$) or continuously replenished \citep{caldiroli2021irradiation,teske2025thick}.

Interestingly, TOI-4616\,b occupies a region of parameter space comparable to, or more extreme than,
TOI-561\,b, for which JWST observations have recently revealed evidence for a substantial atmosphere \citep{teske2025thick}. This suggests that the empirical shoreline is not a strict boundary and highlights TOI-4616\,b as a stringent test of atmospheric survival under extreme irradiation.

\subsection{Atmospheric characterization prospects}

We evaluated the suitability of TOI-4616,b for atmospheric characterization using the Transmission Spectroscopy Metric (TSM) and Emission Spectroscopy Metric (ESM) defined by \citet{kempton2018framework}. Using posterior samples from our joint fit, including contamination corrections, we obtain
$\mathrm{TSM}=21.4$ and $\mathrm{ESM}=2.9$.

\begin{figure}
   \centering
   \includegraphics[width=\columnwidth]{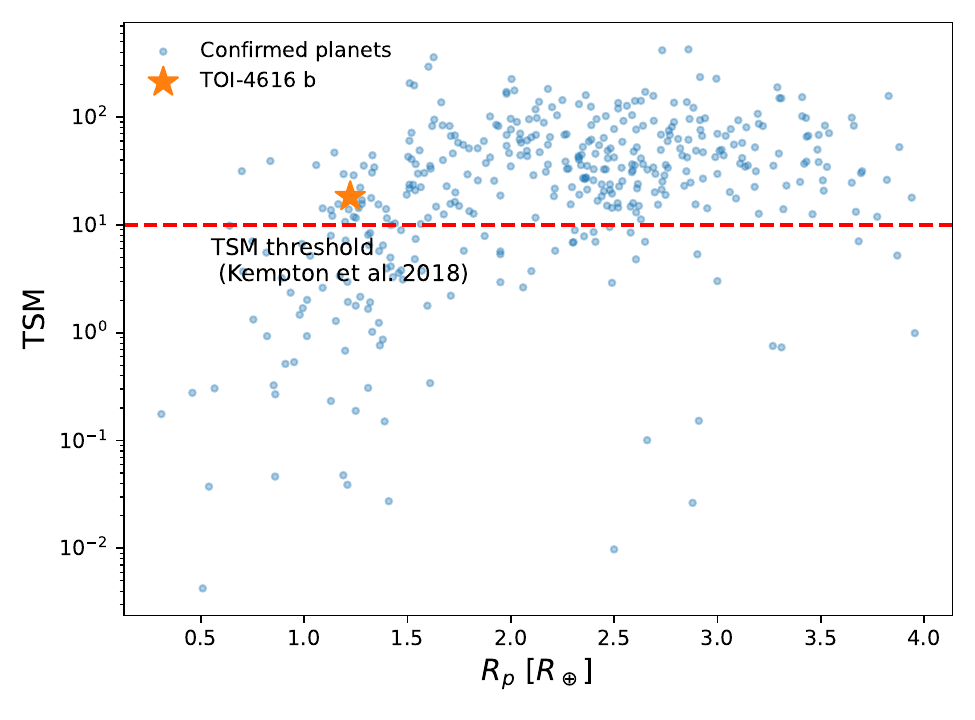}
      \caption{TSM as a function of planetary radius for confirmed transiting planets orbiting M dwarfs. Blue points show the known population compiled from the NASA Exoplanet Archive, accessed on UTC 2026 February 2. The orange star marks TOI-4616\,b. The planet lies above the nominal TSM threshold for transmission spectroscopy of terrestrial planets, indicating that atmospheric features could be detectable with JWST if a secondary atmosphere is present.}
         \label{Fig:TSM}
\end{figure}

The transmission spectroscopy metric (TSM; \citealt{kempton2018framework}) lies above the recommended threshold for atmospheric characterization of terrestrial planets ($\mathrm{TSM} > 10$ for $R_p < 1.5\,R_\oplus$), indicating that transmission spectroscopy with JWST could potentially detect atmospheric features if TOI-4616\,b retains a secondary atmosphere (Fig. \ref{Fig:TSM}). In contrast, the emission spectroscopy metric (ESM) falls below the recommended threshold for thermal emission detection ($\mathrm{ESM} > 7.5$), suggesting that secondary-eclipse observations are unlikely to provide strong constraints on the planet's atmosphere.

TOI-4616,b therefore emerges as one of the few Earth-sized planets orbiting mid-M dwarfs that remains accessible to JWST transmission spectroscopy. Its combination of extreme irradiation, well-constrained stellar properties, and observational accessibility makes it a valuable benchmark for testing models of atmospheric loss, retention, and replenishment in the terrestrial regime.

Continued photometric and spectroscopic monitoring will refine the ephemeris, enable a future mass measurement, and search for additional companions, further strengthening its role as a reference system for highly irradiated rocky planets.

\section{Acknowledgements}

Some of the observations presented in this paper were carried out at the Observatorio Astronómico Nacional on the Sierra de San Pedro Mártir (OAN-SPM), Baja California, México. This work makes use of observations from the LCOGT network; part of the telescope time was granted by NOIRLab through the Mid-Scale Innovations Program (MSIP), funded by the NSF. This research also made use of the Exoplanet Follow-up Observation Program (ExoFOP-TESS; DOI: 10.26134/ExoFOP5), operated by the California Institute of Technology under contract with NASA’s Exoplanet Exploration Program. Computational work was performed on the UBELIX cluster at the University of Bern, and we thank the high-performance computing team for their support.

We acknowledge support from the Swiss National Science Foundation (SNSF) through the SPIRIT project IZSTZ0\_216537 and through the National Centre for Competence in Research PlanetS. Additional institutional support was provided by the Centre for Space and Habitability (CSH) at the University of Bern. Funding for the TESS mission is provided by NASA’s Science Mission Directorate; KAC and CNW acknowledge support via subaward s3449 from MIT.

YGMC, AK, IPF, and MPM acknowledge partial support from UNAM PAPIIT-IG101224. B.-O.~D. acknowledges support from the Swiss State Secretariat for Education, Research and Innovation (SERI) under contract MB22.00046. HPO acknowledges support from the SNSF through PlanetS grants 51NF40\_182901 and 51NF40\_205606. KB acknowledges support from the European Union (ERC Advanced Grant SUBSTELLAR, GA 101054354). This work also received support from NASA under Agreement 80NSSC21K0593 (“Alien Earths”) and benefited from collaborations within the NExSS research coordination network. Support from the ERC Synergy Grant REVEAL (No.\ 101118581) is also acknowledged.

F. M. acknowledges the financial support from the Agencia Estatal de Investigaci\'{o}n del Ministerio de Ciencia, Innovaci\'{o}n y Universidades (MCIU/AEI) through grant PID2023-152906NA-I00.

The ULiege contribution to SPECULOOS has received funding from the European Research Council (grant 336480/SPECULOOS), the Balzan and Francqui Foundations, the Belgian F.R.S.-FNRS (grant T.0109.20), the University of Liège, and the Wallonia-Brussels Federation. M.~Gillon is F.R.S.-FNRS Research Director. The Wallonian Region is thanked for funding the SPECULOOS-North/Artemis telescope. J.d.W. and MIT gratefully acknowledge financial support from the Heising-Simons Foundation, Dr. and Mrs. Colin Masson and Dr. Peter A. Gilman for Artemis, the first telescope of the SPECULOOS network situated in Tenerife, Spain.

This article is based on observations made with the MuSCAT2 instrument, developed by ABC, at Telescopio Carlos S\'{a}nchez operated on the island of Tenerife by the IAC in the Spanish Observatorio del Teide. This work is partly financed by the Spanish Ministry of Economics and Competitiveness through grants PGC2018-098153-B-C31.
This work is partly supported by JSPS KAKENHI Grant Numbers JP24H00017, JP24K00689, and JSPS Bilateral Program Number JPJSBP120249910.

This paper is based on observations made with the MuSCAT instruments developed by the Astrobiology Center (ABC), the University of Tokyo, and LCOGT. MuSCAT3 was developed with support from JSPS KAKENHI (JP18H05439) and JST PRESTO (JPMJPR1775). MuSCAT2 observations were obtained at the Telescopio Carlos Sánchez operated by the IAC at the Observatorio del Teide. Additional support was provided by MCIN/AEI/ERDF projects PID2021-125627OB-C32 and PID2024-158486OB-C32, by the Severo Ochoa Centre of Excellence program, and by JSPS KAKENHI grants JP24H00017 and JP24K00689. F.M. acknowledges support from MCIU/AEI grant PID2023-152906NA-I00.

Observations at the Infrared Telescope Facility were conducted as Visiting Astronomers; the IRTF is operated by the University of Hawai‘i under contract 80HQTR24DA010 with NASA.

\section*{Data availability}

The data underlying this article are publicly available. TESS photometry is available from the Mikulski Archive for Space Telescopes (MAST). Ground-based photometry is available via the Exoplanet Follow-up Observing Program (ExoFOP–TESS). Reduced light curves and analysis scripts used in this work are available from the corresponding author upon reasonable request.





\bibliographystyle{mnras}
\bibliography{example} 


\appendix

\FloatBarrier
\section{Pair-wise \texorpdfstring{$R_p/R_\star$}{Rp/R*} difference distributions between datasets}
We present the distribution of radius ratio differences between datasets, and show how these deviate from zero.

\begin{figure*}
    \includegraphics[width=\textwidth,height=0.8\textheight,keepaspectratio]{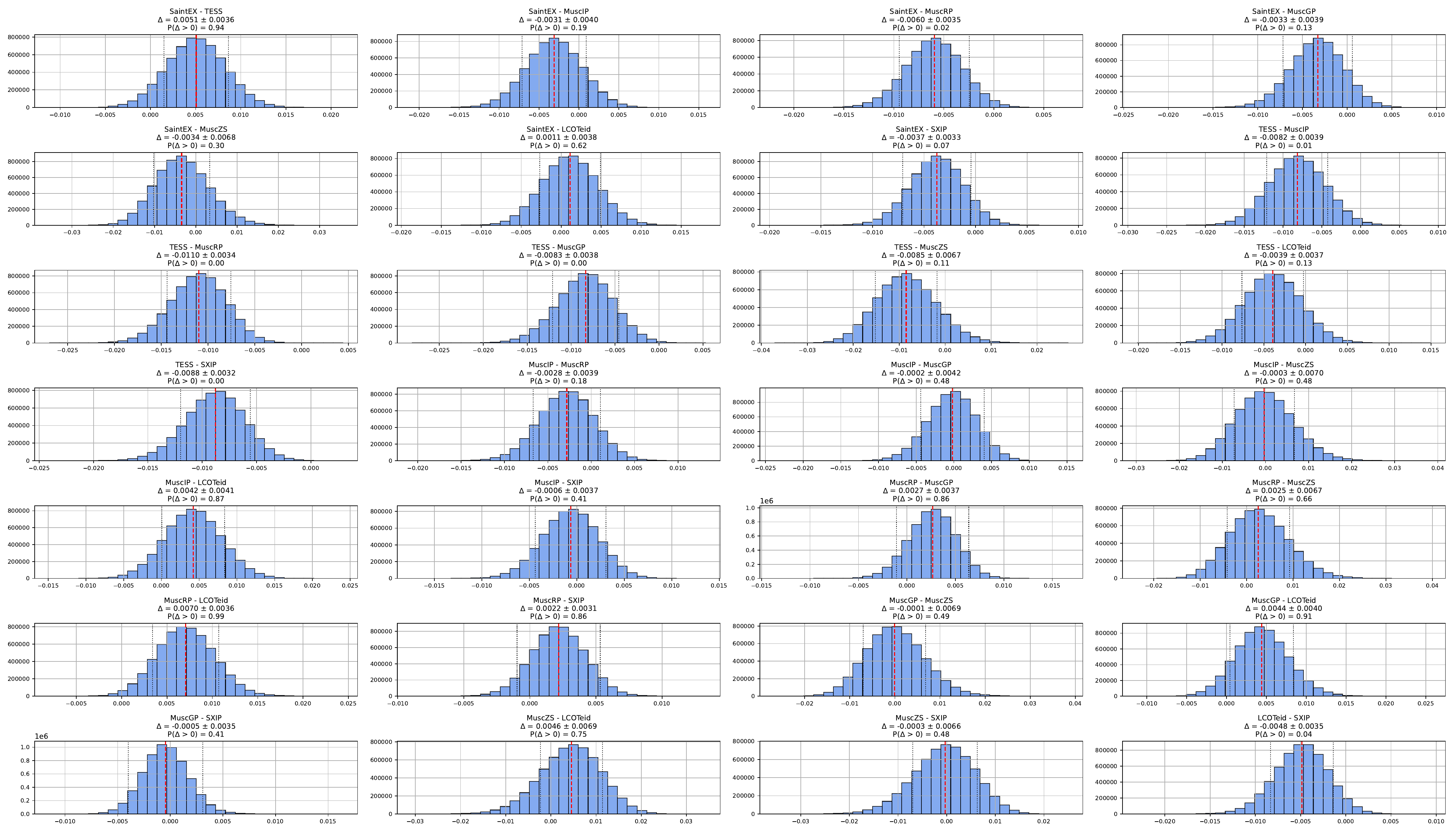}
   \caption{The plots above are the pairwise distributions of the difference in the planet-star radius ratios for all the light curves in the joint fit.
   The broken red vertical lines indicate the median of each distribution. These provide a picture of how much the median differs from zero.
   These plots will indicate a measure of the robustness in the planet-to-star radius ratio calculations.
   We also show the error propagation in the fits, indicating that the median in each distribution lies within the error bars.}
              \label{Fig: ROR1_diff_plots}%
\end{figure*} 

\FloatBarrier
\section{Statistical significance of \texorpdfstring{$R_p/R_\star$}{Rp/R*} differences}

\begin{table*}
\centering
\caption{Bayesian classification of $R_p/R_*$ differences for 28 instrument pairs}
\label{tab:bayesian_ror_comparison}
\begin{tabular}{lrrrrrl}
\toprule
Pair & $\Delta$ & $\sigma$ & $P(\Delta > 0)$ & $Z_\mathrm{Freq}$ & $Z_\mathrm{Bayes}$ & Significance Class \\
\midrule
SAINT-EX - TESS     &  0.0050 & 0.0034 & 0.950 & 1.7353 &  0.9900 & Significant ($\Delta > 0$) \\
SAINT-EX - MuscatIP &  0.0030 & 0.0037 & 0.9190 & 1.0541 &  0.8380 & Not significant \\
SAINT-EX - MuscatRP &  0.0058 & 0.0032 & 0.9910 & 1.8125 &  0.9820 & Significant ($\Delta > 0$) \\
SAINT-EX - MuscatGP &  0.0031 & 0.0034 & 0.9140 & 0.9118 &  0.8280 & Not significant \\
SAINT-EX - LCO+Teid &  0.0010 & 0.0026 & 0.6240 & 0.3846 &  0.2480 & Not significant \\
SAINT-EX - SXIP     &  0.0030 & 0.0030 & 0.9270 & 1.0000 &  0.8540 & Not significant \\
SAINT-EX - MuscatZs     &  -0.0035 & 0.0068 & 0.3000 & 0.515 &  0.5240 & Not significant \\
TESS - MuscatIP    & -0.0060 & 0.0037 & 0.0100 & 1.6216 & -0.9800 & Significant ($\Delta < 0$) \\
TESS - MuscatRP    & -0.0109 & 0.0031 & 0.0000 & 2.7949 & -1.0000 & Significant ($\Delta < 0$) \\
TESS - MuscatGP    & -0.0048 & 0.0034 & 0.0100 & 1.4118 & -0.9800 & Significant ($\Delta < 0$) \\
TESS - LCO+Teid    & -0.0040 & 0.0026 & 0.1120 & 1.5385 & -0.7760 & Not significant \\
TESS - SXIP        & -0.0067 & 0.0029 & 0.0000 & 2.3103 & -1.0000 & Significant ($\Delta < 0$) \\
TESS - MuscatZs    & -0.0005 & 0.0058 & 0.1100 & 0.0862 & -0.7800 & Not significant \\
MuscatIP - MuscatRP&  0.0020 & 0.0035 & 0.1800 & 0.5714 & -0.6400 & Not significant \\
MuscatIP - MuscatGP&  0.0001 & 0.0037 & 0.4900 & 0.0270 & -0.0200 & Not significant \\
MuscatIP - LCO+Teid&  0.0040 & 0.0039 & 0.8600 & 1.2564&  0.7200 & Not significant \\
MuscatIP - MuscatZS& -0.0005 & 0.0070 & 0.4700 & 0.07142& -0.0600 & Not significant \\ 
MuscatIP - SXIP    & -0.0026 & 0.0034 & 0.4100 & 0.7647 & -0.1800 & Not significant \\
MuscatRP - MuscatGP&  0.0028 & 0.0031 & 0.8800 & 0.9032 &  0.7600 & Not significant \\
MuscatRP - MuscatZS& 0.0024  & 0.0067 & 0.4700  & 0.36 & -0.075  &  Not significant \\
MuscatRP - LCO+Teid&  0.0069 & 0.0034 & 0.6600 & 2.0294 &  0.3200 & Not significant \\
MuscatRP - SXIP    & -0.0022 & 0.0027 & 0.3420 & 0.8148 & -0.3160 & Not significant \\
MuscatGP - MuscatZs&  0.0040 & 0.0068 & 0.4700 & 0.5882 & -0.0600 & Not significant \\
MuscatGP - SXIP    & -0.0006 & 0.0029 & 0.3800 & 0.2069 & -0.2400 & Not significant \\
MuscatGP - LCO+Teid&  0.0041 & 0.0036 & 0.9100 & 1.1111 &  0.8200 & Not significant \\
MuscatZs - LCO+Teid&  0.0045 & 0.0070 & 0.7500 & 0.0714 &  0.5000 & Not significant \\
MuscatZs - SXIP    & -0.0045 & 0.0096 & 0.4900 & 0.4688 & -0.0200 & Not significant \\
LCO+Teid - SXIP    & -0.0040 & 0.0032 & 0.0900 & 1.2500 & -0.8200 & Not significant \\
\bottomrule
\end{tabular}
\end{table*}

\FloatBarrier
\section{Corner Plots}

\begin{figure*}
    \includegraphics[width=\textwidth,
                   height=0.9\textheight, keepaspectratio]{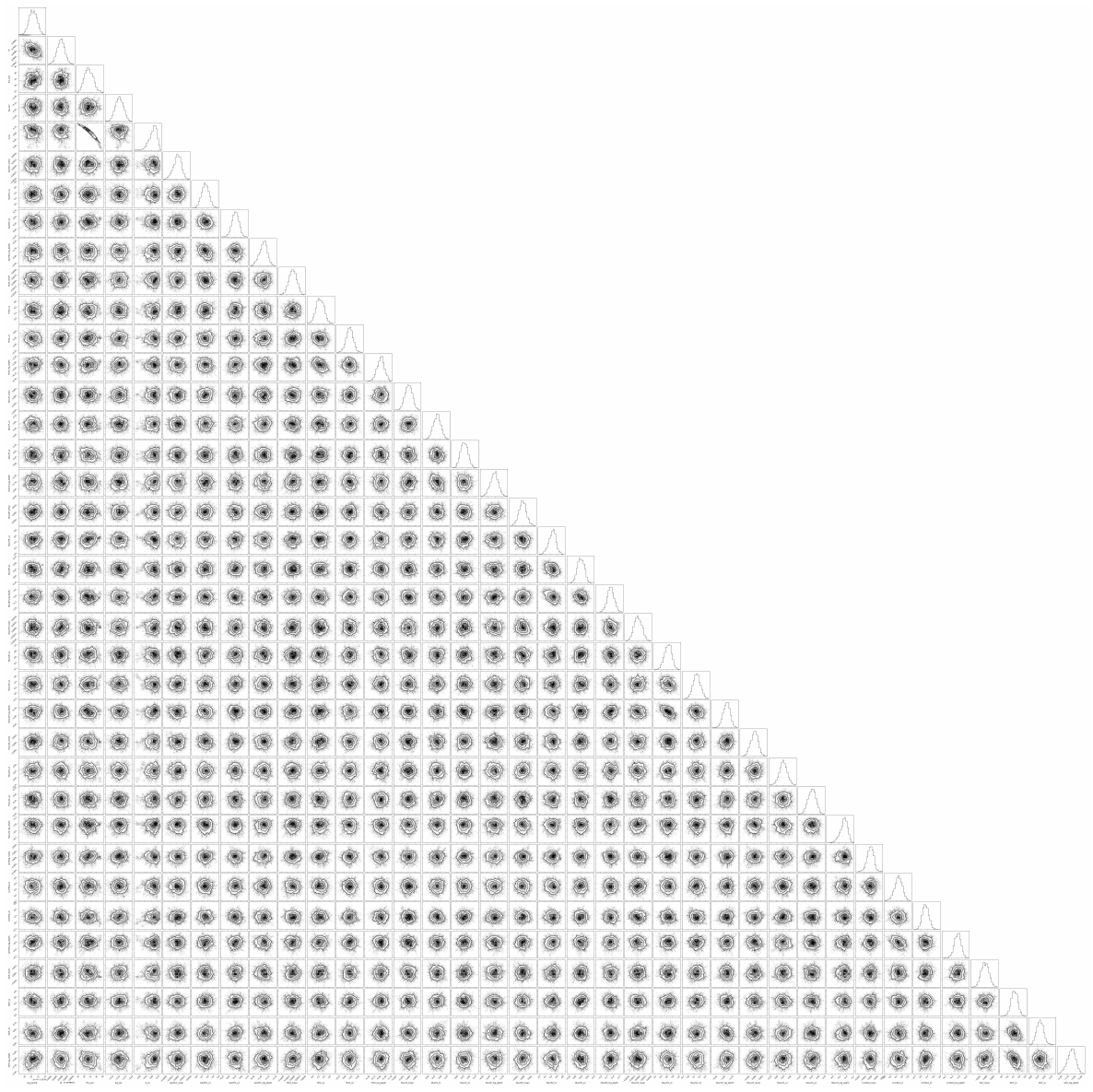}
   \caption{Corner plots for datasets from four instruments (TESS, SAINT-EX, LCO-Teid, and Muscat3- four filters) in five different filters (I+Z, R', G', I', and Z-short). The first correlation in the upper left corner of the plot is that between the mid-transit time and the log of the period. To prevent a very strong correlation between the period and mid-transit time, we placed a prior at the mid point of each dataset. The Calculations that generated these corner plots were carried out on the University of Bern Linux Cluster (UBELIX)}
              \label{cornerSXTESSMuscat}%
\end{figure*} 

\FloatBarrier
\section{Instellation and Equilibrium Temperature}

\begin{figure}
   \centering
   \includegraphics[width=\columnwidth, height=0.4\textheight, keepaspectratio]{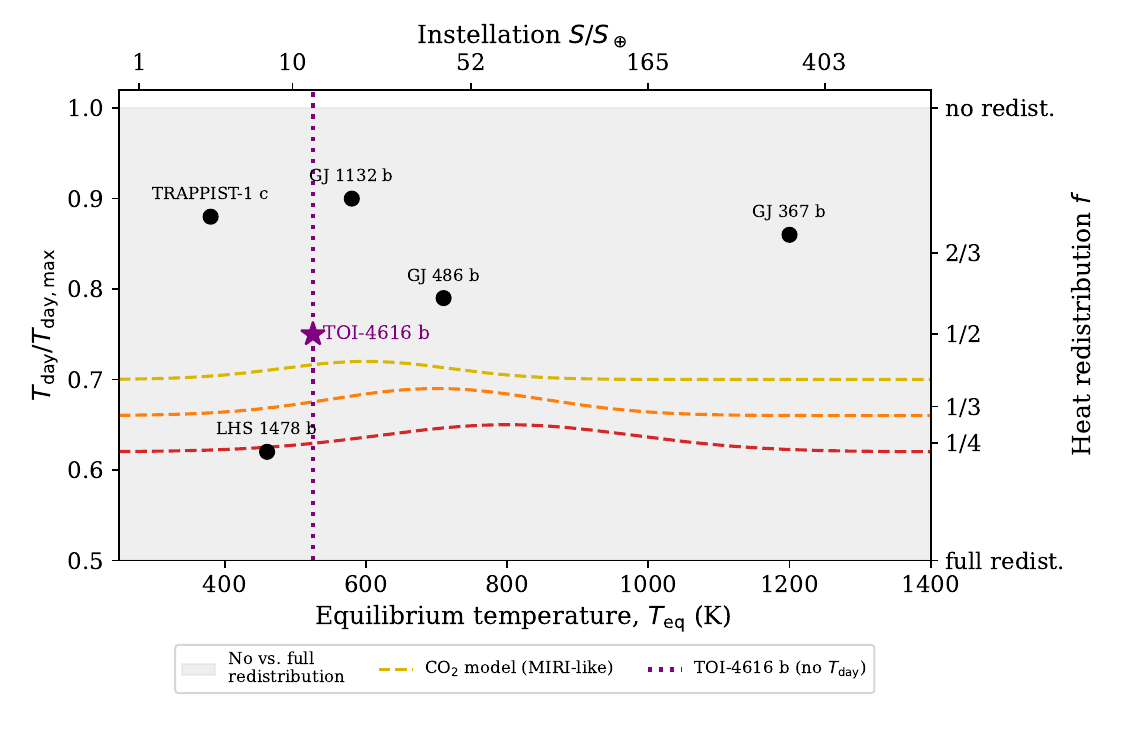}
      \caption{Equilibrium temperature and irradiation diagram illustrating atmospheric heat redistribution regimes for short-period exoplanets. The left axis shows the ratio of dayside temperature to the maximum equilibrium temperature, $T_{\rm day}/T_{\rm eq,max}$, while the right axis gives the corresponding heat redistribution factor $f$. Coloured curves represent theoretical circulation models spanning full to negligible heat redistribution. Black symbols indicate benchmark planets with measured thermal properties. The position of TOI-4616\,b (purple star) is shown based on its equilibrium temperature assuming zero Bond albedo. The diagram illustrates the expected range of day–night heat redistribution efficiencies predicted by atmospheric circulation models. and efficient heat redistribution. No direct constraint on the dayside temperature is currently available.
}
         \label{Fig: JWST_pers}
\end{figure}

\FloatBarrier

\section{Muscat2 Data}
\begin{figure*}
    \centering
    \includegraphics[width=\textwidth]{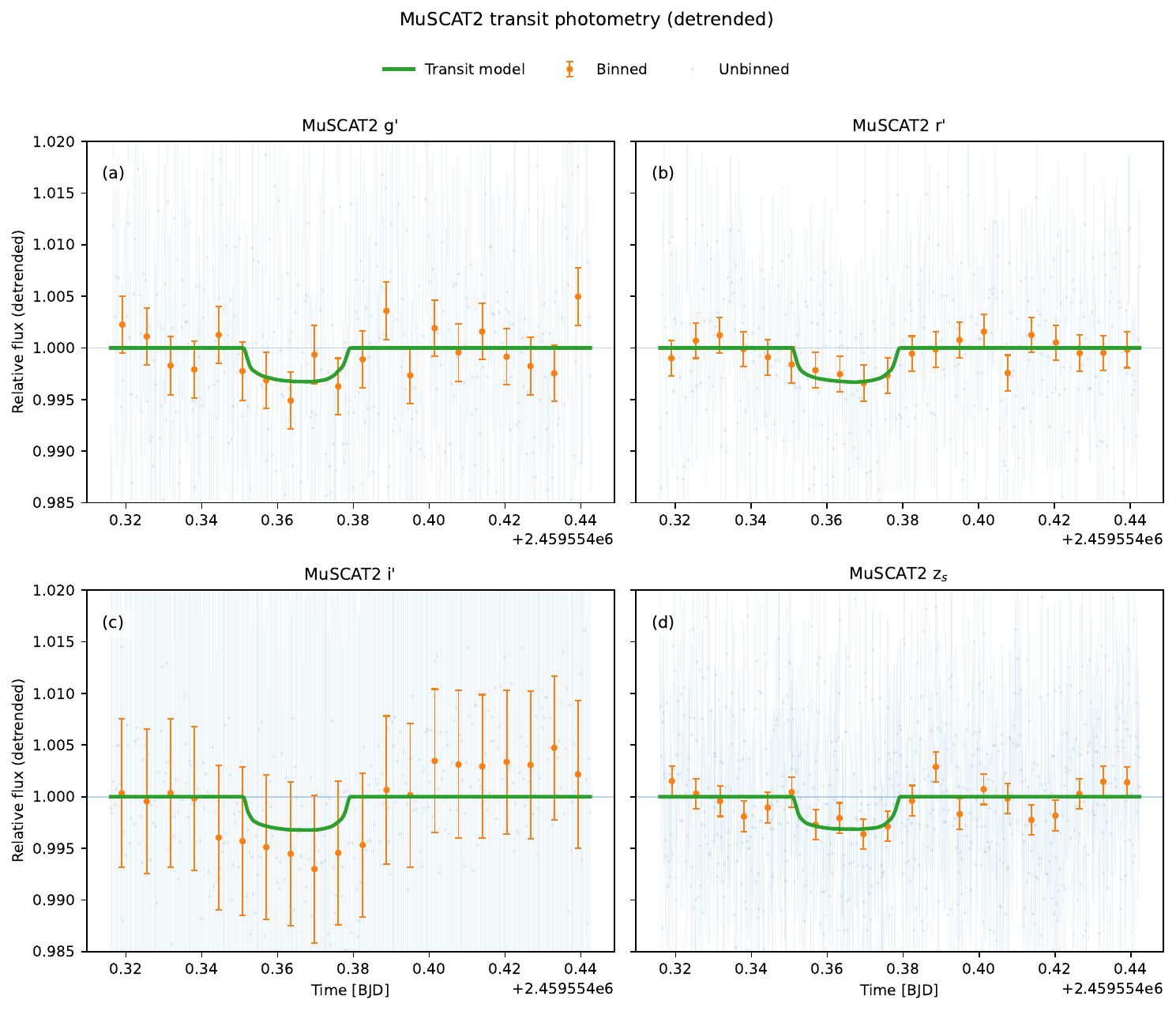}
    \caption{
        MuSCAT2 $g'$, $r'$, $i'$ and $z_\mathrm{s}$ light curves, detrended with
        the quasi–periodic Gaussian Process model from the joint fit.
        Blue points show the raw detrended photometry, orange points are binned
        in time, and the green line indicates the best-fitting transit model.
        Although the individual MuSCAT2 light curves are noisy, the GP
        detrending recovers a coherent transit depth in each band that is
        consistent with the global solution from TESS and higher-S/N
        ground-based instruments, indicating that MuSCAT2 provides a useful
        consistency check on the planetary interpretation.
    }
    \label{fig:muscat2_gp_detrended}
    \label{Fig:MuSCAT2}
\end{figure*}
 
\FloatBarrier
\section{Transit timing analysis}
\label{app:TTV}

Transit Timing Variations (TTVs) can reveal additional non-transiting companions through deviations from a strictly linear ephemeris. To search for such signals in TOI-4616,b, we measured individual mid-transit times for both the TESS and ground-based light curves.

For each transit epoch, we performed an independent fit in which the transit shape parameters (depth, duration, and impact parameter) were fixed to the values obtained from the global joint solution. Only the mid-transit time $T_{0,i}$ and a local baseline model were allowed to vary. The observed-minus-calculated residuals were then computed as

\begin{equation}
(O-C)i = T{0,i} - \left(T_0 + N_i P\right),
\end{equation}

where $T_0$ and $P$ are the reference epoch and orbital period from the global fit, and $N_i$ is the integer transit number.

Figure~\ref{fig} presents the TESS-only timing residuals. The $(O-C)$ values scatter around zero and are consistent with the uncertainties of the individual timing measurements. No coherent periodic signal or systematic drift relative to a linear ephemeris is detected within the available time baseline.

The same procedure was applied to the ground-based transits (SAINT-EX and other facilities), using only fully covered events. The resulting ground-based timing residuals, shown in Fig.~\ref{fig}, are likewise consistent with the linear ephemeris within their larger uncertainties.

We therefore find no statistically significant evidence for transit timing variations in TOI-4616,b. While low-amplitude perturbations or long-period dynamical effects cannot be excluded below the current timing precision, the available data do not indicate the presence of additional planets detectable through TTVs.

\begin{figure}
\centering
\includegraphics[width=\linewidth]{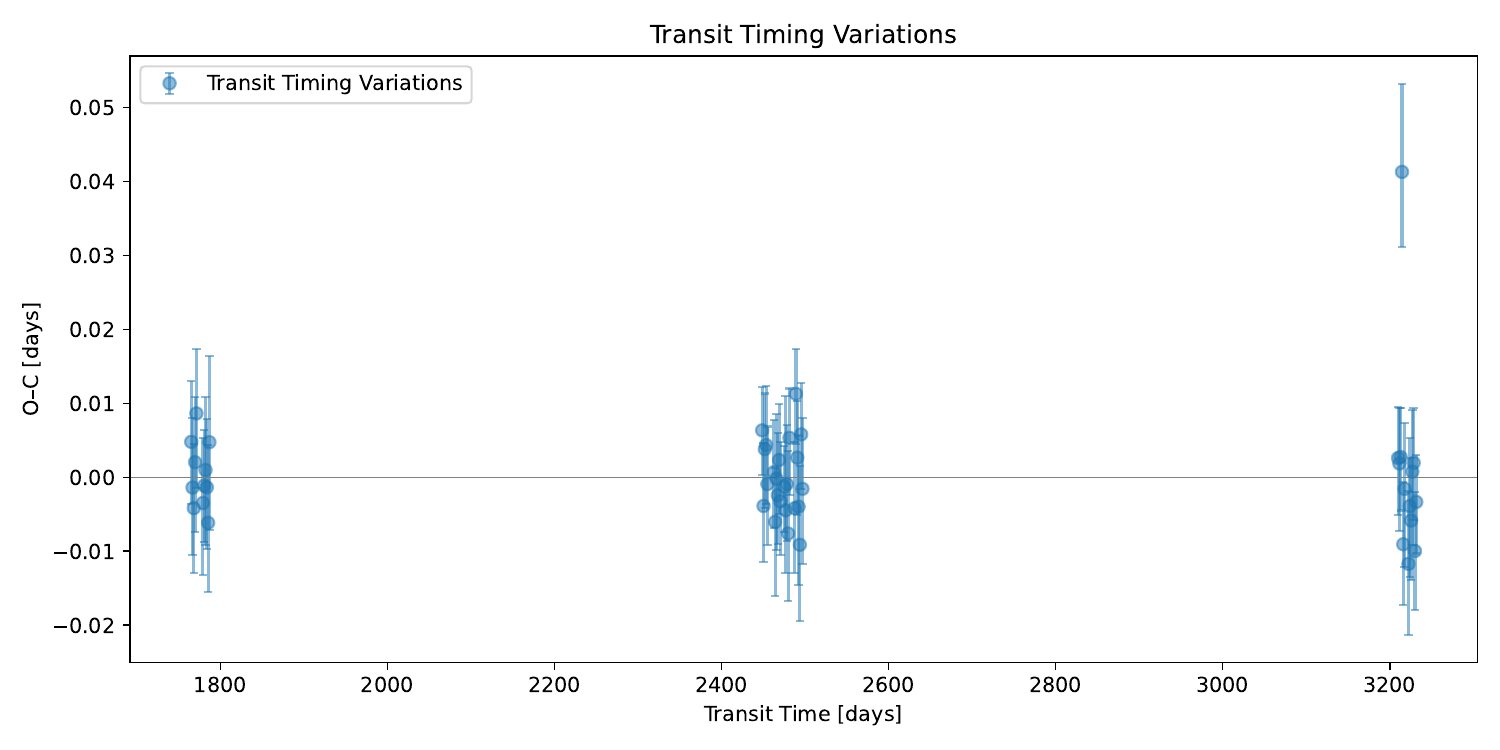}
\caption{TESS-only transit timing residuals for TOI-4616,b relative to the linear ephemeris from the global fit. Error bars represent the uncertainties from the individual timing fits. The residuals are consistent with zero within uncertainties.}
\label{fig}
\end{figure}

\begin{figure}
\centering
\includegraphics[width=\linewidth]{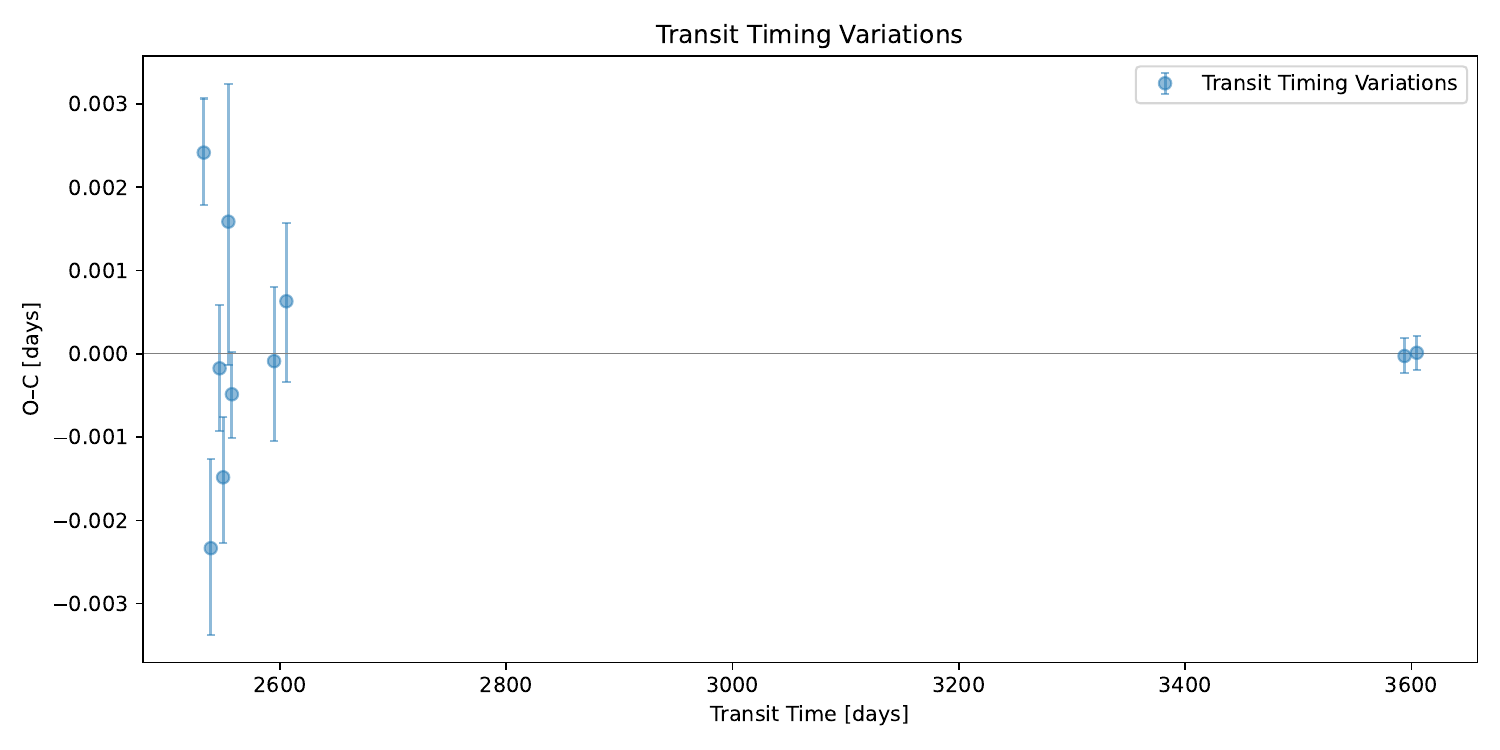}
\caption{Ground-based transit timing residuals for TOI-4616,b derived from individual event fits. The timings are consistent with the linear ephemeris within the larger uncertainties of the ground-based data.}
\label{fig}
\end{figure}


\bsp	
\label{lastpage}

\end{document}